\documentclass[12pt,english]{article}
\usepackage{rotating}
\usepackage{latexsym}
\usepackage{graphicx}
\usepackage{psfrag}
\usepackage{amsmath,amssymb}
\makeatletter
\def\section{\@startsection {section}{1}{\z@}{-3.5ex plus -1ex minus
 -.2ex}{2.3ex plus .2ex}{\large\bf}}
\def\subsection{\@startsection{subsection}{2}{\z@}{-3.25ex plus -1ex
minus -.2ex}{1.5ex plus .2ex}{\normalsize\bf}}
\makeatother
\makeatletter

\@addtoreset{equation}{section}

\makeatother

\def\Dslash{\hspace{3pt}\raisebox{1pt}{$\slash$} \hspace{-9pt} D}

\def\bea{\begin{eqnarray}} \def\eea{\end{eqnarray}}
\def\be{\begin{equation}} \def\ee{\end{equation}} \def\nn{\nonumber}
\def\a{& \hspace{-7pt}}  \def\Z{{\bf Z}}

\def\mc{\mathcal}

\def\tl{\tilde}

\def\s{\hspace{1pt}}

\newcommand{\promille}{%
  \relax\ifmmode\promillezeichen
        \else\leavevmode\(\mathsurround=0pt\promillezeichen\)\fi}
\newcommand{\promillezeichen}{%
  \kern-.05em%
  \raise.5ex\hbox{\the\scriptfont0 0}%
  \kern-.15em/\kern-.15em%
  \lower.25ex\hbox{\the\scriptfont0 00}}

\begin{document}

\thispagestyle{empty}

\begin{center}
\hfill SISSA-77/2005/EP \\
\hfill UAB-FT-590 \\

\begin{center}

\vspace{1.7cm}

{\LARGE\bf A Model of Electroweak Symmetry \\[3mm]
Breaking from a Fifth Dimension}

\end{center}

\vspace{1.4cm}

{\bf Giuliano Panico$^{a}$, Marco Serone$^{a}$ and Andrea Wulzer$^{a,b}$}\\

\vspace{1.2cm}

${}^a\!\!$
{\em ISAS-SISSA and INFN, Via Beirut 2-4, I-34013 Trieste, Italy}

\vspace{.3cm}

${}^b\!\!$
{\em { IFAE, Universitat Aut{\`o}noma de Barcelona,
08193 Bellaterra, Barcelona}}

\end{center}

\vspace{0.8cm}

\centerline{\bf Abstract}
\vspace{2 mm}
\begin{quote}\small

We reconsider the idea of identifying the Higgs field as the internal component
of a gauge field in the flat space $R^4\times S^1/\Z_2$,
by relaxing the constraint of having unbroken SO(4,1) Lorentz symmetry in the bulk.
In this way, we show that the main common problems of previous models of this sort,
namely the prediction of a too light Higgs and top mass, as well as of a too low
compactification scale, are all solved. We mainly focus our attention on a
previously constructed model.
We show how, with few minor modifications and  by relaxing the requirement
of SO(4,1) symmetry, a potentially realistic model can be obtained with a moderate
tuning in the parameter space of the theory. In this model,
the Higgs potential is stabilized and the hierarchy of fermion masses explained.

\end{quote}

\vfill

\newpage

\section{Introduction}

The Electroweak Symmetry Breaking (EWSB) mechanism and the hierarchy of fermion masses
are among the most obscure aspects of the Standard Model (SM).
The minimal set-up of a single doublet scalar field (the SM Higgs field) which drives the EWSB is
affected by a stability problem at the quantum level, since the Higgs
mass term is quadratically sensitive to the scale of new physics.
In the SM, moreover, the observed fermion masses are obtained by an unnatural choice
of Yukawa couplings. Even leaving aside the three neutrinos, their values range from $\sim 10^{-5}-10^{-6}$
for the electron up to $\sim 1$ for the top quark.

Looking for alternative theories in which these problems are solved
has been one of the main guidelines for new ideas and models beyond the SM.
Supersymmetry (SUSY) is certainly the most interesting and well motivated possibility.
It predicts the unification of gauge couplings, it
naturally incorporates a good candidate to explain the dark matter abundance in the universe
and, if assumed to be broken at energy scales $\sim$ few TeV, it can also give rise
to a natural EWSB. Last, but not least, it is a weakly coupled theory.
The simplest model of this sort is the Minimal
Supersymmetric Standard Model (MSSM). Despite the above important
positive aspects, superparticles have not been discovered yet and, in fact,
most of the parameter space of the MSSM is already experimentally ruled out, resulting
in an unwanted fine-tuning in the model. Moreover, the MSSM does not provide
a sensible explanation for the hierarchy of SM fermion masses.
This motivates the quest for other ideas and models, also alternative to SUSY,
which can explain the stability of the EWSB scale and the hierarchy of fermion masses.

Models where the Higgs field is identified with the
internal component of a gauge field in TeV--sized extra dimensions \cite{Antoniadis:1990ew}
(also known as models with gauge-Higgs unification) are an example of this sort
\cite{GHU,burdman} (see \cite{early} for earlier references and \cite{Serone:2005ds} for a brief overview).
The higher dimensional gauge symmetry, rather than SUSY, provides the stabilization
of the Higgs mass term. Consequently, the quadratic divergencies
in the Higgs mass due to the SM particles are cancelled by states with the same statistic,
and not opposite as in SUSY. This is analogous to what typically happens in little Higgs models \cite{LH},
which indeed arose from the deconstructed version of gauge-Higgs unification models \cite{Arkani-Hamed:2001nc}.
The five-dimensional (5D) case, with one extra dimension, is the simplest one and also
the one which seems phenomenologically more appealing.
It is by now clear how to embed the SM fermions and to break the flavour symmetry in such
framework, despite the fact that the Yukawa couplings are gauge
couplings: one can either put the SM fermions on the boundaries and couple
them to massive bulk fermions \cite{Scrucca:2003ra,csaki}
or one can identify the SM fields as the (chiral)
zero modes of bulk fermions with jumping mass terms \cite{Gross-Gher,burdman}.
In both cases one ends up with a concrete realization of the idea of getting
small Yukawa couplings by means of exponentially small overlaps of
wave functions in the internal space \cite{Arkani-Hamed:1999dc}.
Models defined in flat space seem to have common drawbacks. 
Namely, one obtains too low Higgs, top and compactification masses.
To solve these problems, one has to find a gauge-invariant way
to increase the (gauge) couplings of the Higgs with the bulk fermions.
Two known possibilities are the introduction of large
localized gauge kinetic terms \cite{Scrucca:2003ra} and
warped compactification \cite{warped}.
In both cases, however, the bulk wave functions are distorted in a non-trivial
way, resulting in potentially too large
deviations from the SM coming from the Electroweak Precision Tests (EWPT)
and the universality of gauge couplings.
Implementing a custodial symmetry improves the situation, but
some fine tuning is still necessary to get viable models. An interesting
proposal along this direction has been provided in \cite{Agashe:2004rs}.

In this paper we propose a different approach to get a potentially realistic model
with gauge-Higgs unification in flat space.
The essential ingredient which we advocate
is an {\it explicit tree--level breaking of the Lorentz {\rm SO(4,1)} symmetry}.
More precisely, we notice that another possible way to increase the couplings between the Higgs field and the
fermions in a 5D gauge-invariant way is achieved by breaking the SO(4,1)/SO(3,1) symmetry
(so that the usual SO(3,1) Lorentz symmetry is unbroken),
which is the one
 that obliges us to couple the fermions in the same way to
the gauge bosons and to the Higgs field.
In light of this symmetry breaking, we reconsider the minimal 5D model constructed in \cite{Scrucca:2003ra}, to which
we also add a new antiperiodic bulk fermion. The latter state plays an important role to
get a substantial hierarchy between the SM scale and the scale of new physics.
As we will see, such a proposal allows to stabilize
the electroweak scale, explain the hierarchy of fermion masses, get the correct top mass
 and high enough Higgs mass and
compactification scale, then resulting in a potentially very interesting model.\footnote{Another interesting
model with gauge-Higgs unification in flat space is provided in \cite{csaba-giacomo},
where it has been shown that other variants of the model of \cite{Scrucca:2003ra} can also be made realistic.}
As in other models of gauge-Higgs unification, the EWSB is radiatively induced.
The Higgs mass is completely finite at one-loop level. At higher-loops, mainly due to the
Lorentz symmetry breaking, divergencies could be reintroduced, but they would not
spoil the stability of the Higgs potential. The Higgs mass can range from 125 GeV up to 600
GeV (see figure \ref{figmhalphafixmt}) depending on the particular set-up of the model.
The lightest non-standard particle is a colored fermion with mass $M\sim 1-2$ TeV.
Interestingly enough, the neutral component of the
lightest Kaluza-Klein (KK) state of the bulk antiperiodic fermion is a stable weakly
interacting particle with a mass of a few Tev, which is a potential Dark Matter
(DM) candidate, along the lines of \cite{Servant:2002aq}.

The EWPT, the observed suppression of Flavour Changing Neutral Currents (FCNC) and the universality
of the gauge couplings are typically
the most severe tests that any theory which claims to be a realistic extension of the SM
should pass. We do not perform a detailed analysis of these effects, which are left for
future work, but we quantitatively show that our model can pass all these tests.
We argue that FCNC effects should be acceptable, since we estimate
higher derivative operators mediating FCNC to be governed by dimensionless couplings which are naturally
small, particularly for the first two generations.
As in several models based on warped extra dimensions (see {\em e.g.} \cite{Agashe:2003zs,Burd-Caccia,Agashe:2004rs}),
a dangerous and worrisome effect is a deviation from the SM $Zb_L\bar b_L$ coupling.
The resulting bound in our case is about the same one would get from corrections to the four fermion
operators, as computed for theories similar to ours (Higgs and the gauge fields in the bulk, SM fermions
on the brane) \cite{Barbieri:2004qk}. Another important deviation is due to a
custodial breaking mixing between the gauge bosons, which leads to a non-vanishing tree-level
$\Delta \rho$. The resulting bound is, once again, approximately the same as those coming from
$Zb\bar b$ and four fermion operators.
Our theory is compatible with the above phenomenological constraints for a
compactification scale $1/R\gtrsim 5$ TeV. Such values for $1/R$
can be obtained, but at the price of some fine-tuning in the parameter space of the model, which
is hard to quantify in a meaningful way, depending on the prescription used.

{}From a more theoretical point of view, we show that the SO(4,1) Lorentz symmetry
breaking which we advocate can have a natural origin as a spontaneous breaking
induced by a Scherk-Schwarz \cite{SS} twist on a shift symmetry, which can also be seen
as a constant flux for a four-form field strength. This interpretation indicates
that the Lorentz violation we consider can have a natural origin in a 5D framework.

The structure of the paper is as follows. In section 2 we present our model which, as mentioned,
is mainly based on the model of \cite{Scrucca:2003ra}.
In section 3 we show
the predictions of the model, mostly based on numerical results. In particular, we focus on the
values of the Higgs and top masses, and of the compactification scale $1/R$, since the too
low values of these quantities were the main obstructions in constructing
a realistic model of this sort in 5D. In section 4 we roughly quantify the bounds
imposed on our model by $Zb\bar b$, FCNC and the EWPT. We also point out the difficulty
in giving a solid estimate of the amount of the fine-tuning necessary in our model to pass all these tests.
In section 5 we discuss the possible microscopic
origin of the SO(4,1) Lorentz symmetry breaking as a spontaneous breaking
induced by a Scherk-Schwarz twist. Finally, in section 6 we report our conclusions.

\section{The Model}

The model we consider is mainly based on the one built in \cite{Scrucca:2003ra},
namely a $5D$ gauge theory on the $S^1/\Z_2$ orbifold, with group
$G=SU(3)_c\times SU(3)_w\times U(1)^\prime$. We denote by $g_5\equiv g \sqrt{2\pi R}$ and 
$g_{5}^\prime\equiv g^\prime \sqrt{2\pi R}$
the $SU(3)_w$ and $U(1)^\prime$ gauge couplings respectively. The extra $U(1)^\prime$ and its coupling $g^\prime$
have to be introduced in order to get the correct weak mixing angle.
The  $\Z_2$ orbifold projection is embedded non-trivially in the electroweak $SU(3)_w$
group only, by means of the matrix
\begin{equation}
P = e^{2 \pi i \sqrt{3} t_8} =
\left(
\begin{matrix}
-1 \a 0 \a 0 \cr
 0 \a -1\;\;\, \a 0 \cr
 0 \a  0 \a 1 \cr\end{matrix}\;
\right)\;,\label{Rtwist1}
\end{equation}
where $t_a$ are the standard $SU(3)$ generators,
normalized as ${\rm Tr}\, t_a t_b = 1/2 \delta_{ab}$.
The twist (\ref{Rtwist1}) breaks the electroweak gauge group to
$SU(2) \times U(1)\times U(1)^\prime$ in $4D$.
The massless 4D fields are the gauge bosons in the adjoint of $SU(2)\times U(1)\subset SU(3)$, the gauge field $A_\mu^\prime$
and a charged scalar doublet $H$, the Higgs field, arising from $A_5^a$.
The hypercharge generator $Y$, such that $Y=1/2$ for the Higgs field, is taken to be the linear combination
$Y=\frac1{\sqrt{3}}t_8+t^\prime$ of the $U(1)$ and $U(1)^\prime$ generators.
The gauge field $A_Y$ associated to the hypercharge and its orthonormal combination $A_X$ are
\begin{equation}
A_Y=\frac{g^\prime A^8+\sqrt{3}g A^\prime}{\sqrt{3g^2+{g^\prime}^2}}\,,\;\;\;\;\;
A_X=\frac{\sqrt{3}g A^8-g^\prime A^\prime}{\sqrt{3g^2+{g^\prime}^2}}\,.
\end{equation}
The $U(1)_Y$ coupling $g_Y$ is related to the $4D$ $SU(2)$ and $U(1)^\prime$ couplings $g$ and
$g^\prime$ as
$g_Y=\sqrt{3}g g^\prime/\sqrt{3 g^2+{g^\prime}^2}$.
By suitably choosing $g^\prime$ we can adjust the weak angle to the correct value, according to the relation
\begin{equation}
\sin^2\theta_W=\frac{{g_Y}^2}{g^2+{g_Y}^2}=\frac{3}{4+3g^2/{g^\prime}^2}\,.
\end{equation}
As we will better explain in subsection 2.2, the zero mode of $A_X$ will get a large mass and decouple from the theory,
leaving only its KK excitations. A vacuum expectation value (VEV) for $A_5^a$
induces an additional spontaneous symmetry breaking to $U(1)_{EM}$.
We can take
\begin{equation}
\langle A_5^a \rangle \equiv \frac {2\alpha}{g_5 R}\,\delta^{a7} \,,
\label{vev}
\end{equation}
corresponding to an imaginary VEV for the Higgs field: $\langle H\rangle = 2i\alpha/(g R)$.
The parameter $\alpha\in[0,1/2]$ in Eq.~(\ref{vev}) is a Wilson line phase,
and thus the EWSB in this model is equivalent
to a Wilson line symmetry breaking \cite{hos}.

\subsection{The matter Lagrangian}

Introducing matter fields in this set-up is a non-trivial task.
One possibility is to include
massive 5D bulk fermions and massless localized chiral fermions, with a mixing
between them, so that the matter fields are identified with the lowest KK mass
eigenstates. In this way, Yukawa couplings are exponentially sensitive to the bulk
mass terms, and the observed hierarchy of fermion masses is naturally explained.\footnote{
Note that bulk-brane systems of fermions of this kind
could naturally originate from a single bulk field on a resolved orbifold, along the
lines of \cite{cal}. The bulk-brane spectra considered here are however not compatible
with those found in \cite{cal}.}
Here we focus on the third generation
of quarks (top and bottom), since light quarks and leptons do not significantly
contribute to the Higgs potential.\footnote{See however \cite{csaba-giacomo} for the study of a
different set-up, in which other quarks and leptons can significantly contribute to the Higgs
potential.}
The bulk 5D fermions that have the correct quantum numbers to couple with
the (conjugate) top and with the bottom are respectively  the symmetric (${\bf 6}$) and
fundamental (${\bf 3}$) representations  of $SU(3)_w$, neutral under the $U(1)^\prime$ group.
In addition to that, we also add a symmetric representation, antiperiodic on the covering circle
$S^1$, with $U(1)^\prime$ charge 2/3. We impose a global $U(1)_A$ symmetry
under which only the antiperiodic bulk fermions transform. This symmetry forbids any
mixing between these fields and the localized ones.\footnote{This $U(1)_A$ symmetry, as well as
the $U(1)^\prime$ charge we have chosen for the antiperiodic fermions, are introduced
uniquely to possibly get a viable DM candidate out of these states.} 
Such state was not present in the original model of \cite{Scrucca:2003ra}.

The matter fermion content of this basic construction is more precisely the following.
We introduce a couple of periodic bulk fermions ($\Psi_t$, $\widetilde\Psi_t$)
with opposite $\Z_2$ parities, in the representation $(\bar{\bf 3},{\bf 6})$ and
($\Psi_b$, $\widetilde\Psi_b$) in the $({\bf 3},{\bf 3})$ of $SU(3)_c\times SU(3)_w$ and a
couple of antiperiodic bulk fermions $\Psi_A$ and $\widetilde\Psi_A$
with opposite $\Z_2$ parities, in the $({\bf 1},{\bf 6})$.
All these fermions have unconventional
SO(4,1) Lorentz violating kinetic terms.
At the orbifold fixed points, we have a
left-handed doublet $Q_L=(t_L,b_L)^T$ and two
right-handed fermion singlets $t_R$ and $b_R$ of $SU(2)\times U(1)$.
They are located at $y_1$ and $y_2$, equal to $0$ or $\pi R$,
the two boundaries of the segment.
The parity assignments for the bulk fermions allow for a bulk mass term
$M$ mixing $\Psi$ and $\widetilde \Psi$, as well as boundary
couplings $e_{1,2}$ with mass dimension $1/2$ mixing the
bulk fermions to the boundary fields $Q_L$, $t_R$ and $b_R$.
The matter Lagrangian reads\footnote{The flavour structure of the full model, including all
quarks and leptons, is obtained exactly as in \cite{Scrucca:2003ra}, with the only
difference that now one could introduce an SO(4,1) Lorentz violating matrix $k_{ij}$, which
provides an additional source of flavour mixing. An interesting alternative would be to
introduce a flavour symmetry in the model, along the lines of \cite{Martinelli:2005ix}.}
\begin{eqnarray}
\mathcal{L}_{\rm mat} \a = \a
\sum_{j=t,b,A} \Bigg\{ \overline \Psi_j  \Big[i \Dslash_4 -  k_j D_5 \gamma^5\Big] \Psi_j
+ \overline{\widetilde \Psi}_j \Big[i \Dslash_4 -  \tilde k_j  D_5 \gamma^5\Big] \, \widetilde \Psi_j
 + \Big(\overline \Psi_j M_j \widetilde \Psi_j +\mathrm{h.c.} \Big)\Bigg\} \nn \\
 \a\a  + \delta(y-y_1) \Big[ \overline Q_L \,i \Dslash_4\, Q_L +
\big(e_1^b \overline Q_L \psi_{b} + e_1^t \overline Q_R^c \psi_{t} +\mathrm{h.c.} \big)\Big] \nn \\
\a\a  + \delta(y-y_2) \Big[\overline t_R \,i \Dslash_4\, t_R + \overline b_R \,i \Dslash_4\, b_R
 + \big(e_2^b \overline b_R \chi_{b} + e_2^t \overline t_L^c \chi_{t}+\mathrm{h.c.} \big)\Big]\,,
\label{Lagferm}
\end{eqnarray}
where $\psi_{t,b}$ and $\chi_{t,b}$ are the doublet and singlet $SU(2)$ components
of the bulk fermions $\Psi_{t,b}$. For simplicity, in the following we take $k_j=\tilde k_j$.
The metric is ``mostly minus'' and $(\gamma^5)^2=1$.
All bulk fermion modes are massive and, neglecting the
bulk-to-boundary couplings, their mass spectrum is given by $M_{n,j} =
\pm \sqrt{m_{n,j}^{2} + M_j^2}$, where $m_{n,j}=k_j n/R$.  When the EWSB
induced by (\ref{vev}) is considered, a new basis has to be defined
for the bulk fermion modes in which they have diagonal mass
terms, with a shift in the KK masses $m_{n,j} \rightarrow
m_{n,j}(\alpha)$. The procedure is outlined in the appendix of \cite{Scrucca:2003ra}.

In the following, it will be convenient to take the size $\pi R$ of
the orbifold as reference length scale and use it to define
dimensionless quantities. In particular, it will be useful to
introduce the parameters $\lambda^i = \pi R M_i$ and $\epsilon_i^a =
\sqrt{\pi R/2} e_i^a$.

\subsection{Gauge bosons and anomaly cancellation}

The localized chiral fermions in our model induce gauge and gravitational anomalies
that must be cancelled. As already discussed in \cite{Scrucca:2003ra}, the precise pattern
of anomaly cancellation depends on the position of the localized fermions.
For all distributions of matter, all anomalies can be cancelled, but at the
price of introducing two localized axions (at $y=0$ and at $y=\pi R$) and a
Chern-Simons term with a jumping coefficient \cite{anomaly}, which
has to be introduced anytime the SM anomalies (the ones which do not involve the field $A_X$)
do not locally cancel in the internal direction.
When a Chern-Simons term is needed, couplings which are $\Z_2$ odd cannot be anymore
consistently neglected. For this reason, for simplicity,
we focus in the following on two special set-ups, that we shortly denote $\delta=0$ and $\delta=1$,
in which all the SM anomalies are locally cancelled.
We call $\delta=0$ the set-up in which all SM fermions are located at the same fixed-point (say, at $y=0$).
Among all the various set-ups in which the matter is located in both fixed-points, but in such a way
that the SM anomalies cancel locally, we call $\delta=1$ the ones in which the ($t_L,b_L$) doublet is, say, at $y=0$,
whereas $t_R$ and $b_R$ are at $y=\pi R$, without specifying in detail the location of the other SM fermions.

The anomalies which are left are those involving $A_X$ and can be cancelled by means of a 
4D version of the Green--Schwarz mechanism (GS) \cite{GS}.
One introduces one ($\delta=0$) or two ($\delta=1$) localized axions, transforming non-homogeneously
under the $U(1)_X$ symmetry, with non-invariant 4D Wess--Zumino
couplings that compensate for the one-loop anomaly. In this way all mixed
$SU(3)_c\times SU(2)_L \times U(1)_Y \times U(1)_X$ gauge and
gravitational anomalies can be cancelled. When $\delta=0$, the single axion is eaten by the
gauge field $A_X$, whereas for $\delta=1$ one combination of them is eaten, while the orthogonal
one remains massless. In a suitable gauge, the net effect of the anomaly on the gauge bosons is
the appearance of localized quadratic terms in $A_X$, with a mass term $M_X$ whose natural size is
the cut-off scale of the model. For $M_X\gg 1/R$,
the localized mass terms simply result in an effective change (from Neumann to Dirichlet) of the boundary condition of $A_X$
at the points where they are located. In the limit $M_X\rightarrow \infty$ and for $-\pi R\leq y \leq \pi R$, 
the KK expansion of $A_X$ is given by
\be
A_{X}(x,y)=  \frac1{\sqrt{\pi R}}\sum_{n=0}^{\infty}A_{X,\,n}
(x)\sin\frac{[n+(1+\delta)/2]|y|}{R}\,.
\label{axwf}
\ee
The zero--mode of $A_X$ decouples while its KK tower is still at low energy and can have sizable effects.\footnote{Ref.\cite{Scrucca:2003ra}
overlooked these effects, considering only the mixing of the SM $Z$ boson with the zero mode of $A_X$.}
The remaining gauge bosons are insensitive to the localized mass terms and retain their original
expansion in cosines or sines as given by eq.(\ref{Rtwist1}).

When the EWSB occurs, the diagonal mass eigenstates, taking into account of the localized mass terms
for $A_X$, are the following:
\be
\begin{array}{l@{\hspace{2.5em}}l}
m_n^{(1)}  =  \displaystyle\frac{n+\alpha}{R}\,,  \ \ \ \ 
m_n^{(2)}  = \frac{n+h_\delta(\alpha)}{R}\,, \a n\in [-\infty,\infty]\,, \nn \\
\rule{0pt}{2em}m_n^{(3)}   =  \displaystyle\frac{n}{R}\,, \ \ \ m_n^{(4)}=\frac{n+1}{R}\,, \ \ \ m_n^{(5)} = \frac{n+(1+\delta)/2}{R}\,,  
\a n\in [0,\infty]\,.   
\end{array}
\label{gaugemasses}
\ee
The SM gauge bosons $W$, $Z$ and $\gamma$ are associated to the $n=0$ modes of $m_n^{(1)}$, $m_n^{(2)}$ and $m_n^{(3)}$, 
so that the $W$ mass equals
\be
m_W=\frac{\alpha}{R}\,.
\label{mw}
\ee
The functions $h_\delta(\alpha)$ appearing in eq.(\ref{gaugemasses}) originate from the
localized mass terms of $A_X$. In the limit $M_X\rightarrow\infty$, $m_Z=h_\delta(\alpha)$ 
is defined by the transcendental mass equations
\bea
\sin^2(\pi m_Z R)\a = \a \frac{1}{4\cos^2\theta_W}
\sin^2(2 \pi \alpha)\,, \hspace{3.7cm} \delta=0 \nn \\
\sin^2(\pi m_Z R)\a = \a \frac{1}{\cos^2\theta_W}
\sin^2(\pi \alpha)-\frac{1}{4\cos^4\theta_W} \sin^4(\pi \alpha)\,. \ \ \ \delta=1
\label{mzexact}
\eea
By expanding the sines in eq.(\ref{mzexact}) one finds at leading order the SM relation $m_Z=m_W/\cos\theta_W$, as expected.
Corrections due to the localized mass terms for $A_X$ are however present, so that $\rho\neq 1$. 
We will better quantify such corrections in subsection 4.2.

By 5D gauge symmetry, the Higgs mass vanishes at tree-level and is radiatively induced. It equals
\begin{equation}
m_H^2(\alpha_{min}) = \bigg(\frac{g_4 R}2\bigg)^2 \,
\frac{\partial^2 V}{\partial \alpha^2}\bigg|_{\alpha=\alpha_{min}} \;,
\label{MH}
\end{equation}
with $V(\alpha)$ the (radiatively induced) Higgs effective potential and $\alpha_{min}$ its minimum.

\subsection{Higgs potential and induced fermion masses}

The 5D $SU(3)_w$ gauge symmetry, which is not broken by the Lorentz violating
couplings $k_j$, forbids the appearance of any local Higgs potential in the bulk.
An Higgs potential localized at the orbifold fixed points is also forbidden
by a non-linearly realized symmetry which is left unbroken by the orbifold
boundary conditions. This symmetry acts on the Higgs field components $A_5^a$ ($a=4,5,6,7$)
as \cite{vonGersdorff:2002rg}
\be
A_5^a \rightarrow A_5^a + \partial_5 \xi^a\,.
\label{shift}
\ee
The symmetry (\ref{shift}) is not broken in our model and hence
we expect that the Higgs potential is still radiatively induced by non-local operators and thus
finite. Since the field $A_5$ couples only to the gauge fields and to the bulk
fermions, its potential depends indirectly on the boundary couplings
through diagrams in which the virtual bulk fermions temporarily switch to a virtual boundary
fermion.

The one loop contribution to the potential given by the 5D gauge bosons and ghosts
is easily computed from the explicit form of the KK mass spectrum (\ref{gaugemasses}).
It is given by\footnote{Eq.(\ref{Vgauge}) is 
obtained by replacing $2\alpha\rightarrow h_\delta(\alpha)$ in eq.(29) of \cite{Scrucca:2003ra},
that overlooked the corrections due to the $U(1)_X$ anomaly.} 
\begin{equation}
V_{g}(\alpha) = 2 V_{A}(\alpha) + V_{A}[h_\delta(\alpha)] \,,
\label{Vgauge}
\end{equation}
where
\begin{equation}
V_{A}(\alpha) = -\frac{9}{64 \pi^6 R^4}\sum_{k=1}^\infty \frac{1}{k^5}
\cos (2 k \pi \alpha) \,.
\label{V5gauge1}
\end{equation}

The one loop contribution from a massive 5D fermion with mass $\lambda$ and given $k$ is also
easily found.\footnote{
As far as the contribution of a single fermion $\Psi$ is concerned,
the factor $k$ can be eliminated by a redefinition of the $y$ coordinate, which results
in the following rescaling of the parameters:
\begin{equation}
\left\{
\begin{array}{l}
R \rightarrow R/k\, ,\\
\lambda \rightarrow \lambda/k\, ,\\
\varepsilon_i \rightarrow \varepsilon_i/k\, ,
\end{array}
\right.
\label{rescaling}
\end{equation}
and a rescaling $\Psi\rightarrow \Psi/\sqrt{k}$.
This procedure can be used to derive the bulk fermion contributions in eq.(\ref{Va-xCorr}).
However, the boundary contributions in eqs.(\ref{boundpott}) and (\ref{boundpotb}),
in which two fields with different $k$'s
are involved, cannot be obtained by such simple scaling argument.}
For a pair of modes with charge $q$, one has
\be
V_{\Psi}(q\alpha) = \frac{3 k^4}{8 \pi^6 R^4}\sum_{n=1}^\infty n^{-5}\Big(
1 + 2 n \frac{\lambda}{k} + \frac 43 n^2 \frac{\lambda^{2}}{k^2}\Big) e^{- 2 n \frac{\lambda}{k}}
\cos \Big[2 \pi n q (\alpha+\frac{\eta}{2q})\Big] \,,
\label{Va-xCorr}
\ee
where $\eta=0$ for periodic fermions and $\eta=1$ for antiperiodic fermions.

The full Higgs effective potential is obtained by summing the gauge and fermion
contributions, including also the contributions of the fermion boundary terms.
The explicit formulae for the latter ones 
can be derived exactly as in \cite{Scrucca:2003ra}, modulo the changes due to the SO(4,1)
breaking parameters, and are given by (see \cite{Scrucca:2003ra} for the notation)\footnote{In eq.(25)
of \cite{Scrucca:2003ra} there is a typo
in the last term: $f_\delta(x^u,\alpha)$ should be replaced by $f_\delta(x^u,2\alpha)$.}
\begin{eqnarray}
\!\!\!\! V_{t}(\alpha) \a=\a \frac{-1}{4\pi^6 R^4}
\int_0^\infty \!\!\!\!\! dx\,x^3\,
{\rm ln}\Bigg[\prod_{i=1}^2{\rm Re} \Big[1
+ \delta_{i1} \frac{\epsilon_1^{b\s 2}\!}{k_b x^b} f_0(\frac{x^b}{k_b},0)
+ \delta_{i2} \frac{\epsilon_2^{t\s 2}\!}{2k_t x^t} f_0(\frac{x^t}{k_t},0) \nn \\
\a\!\a \hspace{85pt} +\, \frac{\epsilon_i^{t\s 2}\!}{2^{\delta_{i2}}k_t x^t}
f_0(\frac{x^t}{k_t},2\alpha) \Big]
+ \prod_{i=1}^2{\rm Im} \Big[\frac {\epsilon_i^{t\s 2}\!}{2^{\delta_{i2}}k_t x}
f_\delta(\frac{x^t}{k_t},2\alpha)\Big]\Bigg] , \label{boundpott} \hspace{15pt}\\
\!\!\!\! V_{b}(\alpha) \a=\a \frac{-1}{4\pi^6 R^4}\int_0^\infty \!\!\!\!\! dx\,x^3\,
{\rm ln}\Bigg[\prod_{i=1}^2{\rm Re} \Big[1
+ \frac{\epsilon_i^{b\s 2}\!}{k_b x^b} f_0(\frac{x^b}{k_b},\alpha)
+ \delta_{i1} \frac{\epsilon_1^{t\s 2}\!}{k_t x^t} f_0(\frac{x^t}{k_t},\alpha)
\Big] \nn \\ \a\;\a \hspace{85pt}
+ \prod_{i=1}^2{\rm Im} \Big[\frac {\epsilon_i^{b\s 2}\!}{k_b x}
f_\delta(\frac{x^b}{k_b},\alpha)\Big]\Bigg] \;. \label{boundpotb}
\end{eqnarray}

We find that the presence of antiperiodic fermions is necessary to obtain
small enough values of $\alpha_{min}$. Indeed, as it can be seen
from eq.~(\ref{Va-xCorr}), they permit a partial cancellation
of the leading cosine in the fermion contribution to the potential to be enforced,
then lowering the position of its global minimum
\cite{susy} (see also \cite{Haba}).
Note that for this cancellation to take place a certain correlation
among the parameters, mainly between $k_t$ and $k_A$, is required.
We better quantify it in section 3.
The Higgs mass, however, is generically too low in this set-up for $k_i = 1$.
Higher values of  $k_i$ considerably help in getting higher Higgs masses.
This is particularly clear in the rough approximation in which one neglects the  boundary
contributions (\ref{boundpott}) and (\ref{boundpotb}) (as well as the gauge contribution)
to the Higgs potential, and takes
$k_b=k_t=k_A=k$ and massless 5D bulk fermions: $\lambda_i=0$. In this case, the total Higgs potential is
given by the sum of the bulk contributions of the form (\ref{Va-xCorr}) (with $\lambda=0$).
This is exactly of the same form as the usual SO(4,1) invariant case, except for an overall
$k^4$ factor in front of the potential. According to eq.(\ref{MH}), the Higgs mass
is $k^2$ times the Higgs mass evaluated in the standard case with $k=1$.
As we will see below, the factors $k_i$'s are also crucial to get reasonable top masses.

\begin{figure}
\begin{tabular}{c@{\hspace{2em}}c}
\includegraphics[width=.45\textwidth]{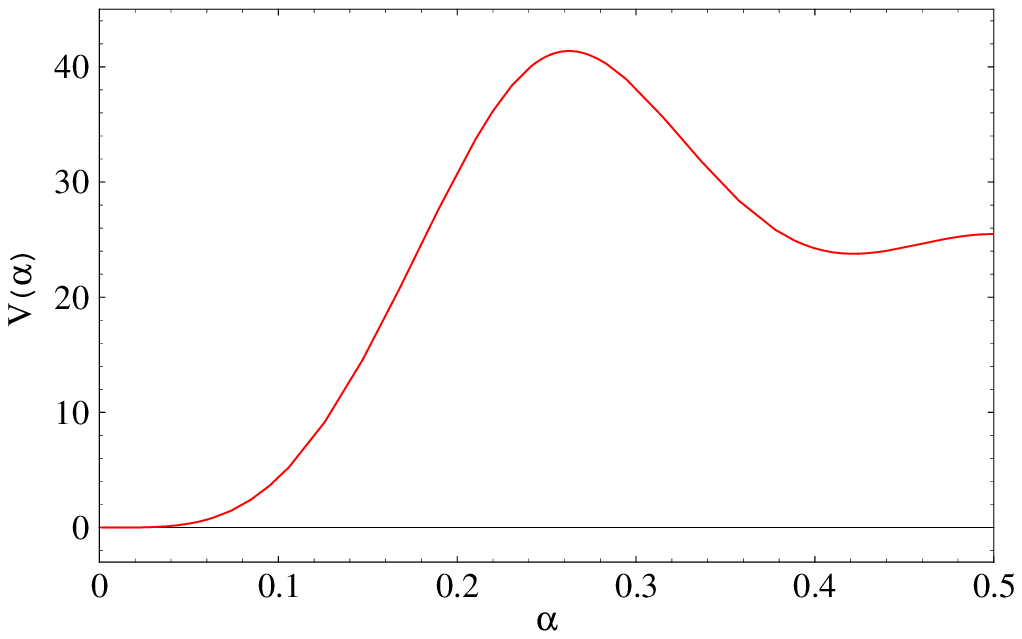}&
\includegraphics[width=.475\textwidth]{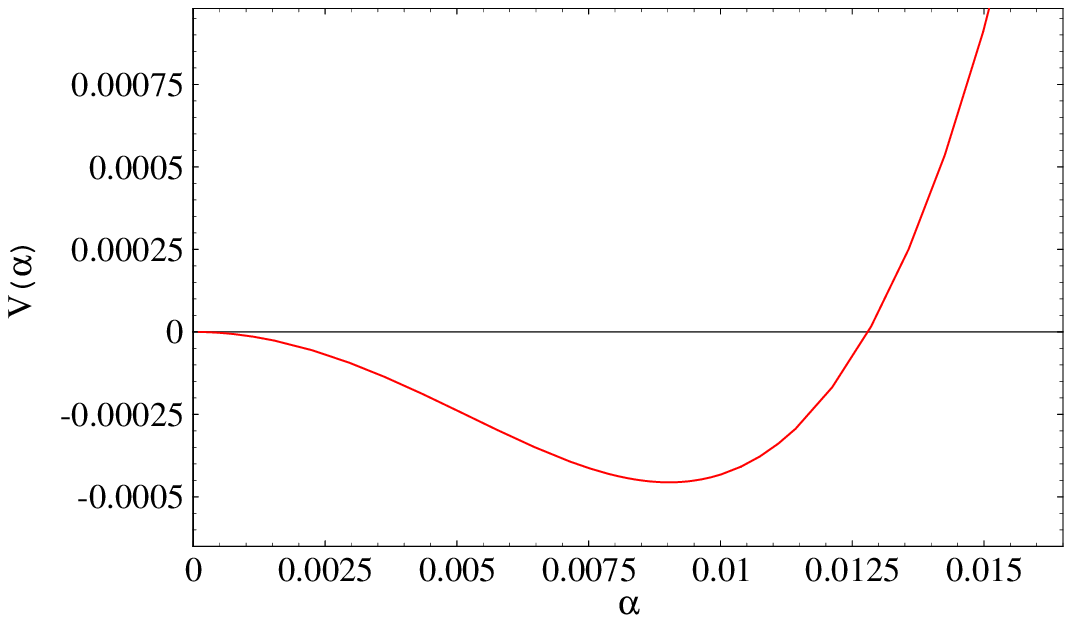}
\end{tabular}
\caption{The Higgs potential in the $\delta=0$ set-up, obtained with input parameters
$\lambda^t = 0.99$, $\lambda^b = 6.9$, $\lambda^A = 0.24$, $k_t = 2.42$, $k_b = 2.26$, $k_A = 3.14$, $\varepsilon_1^t=1.9$,
$\varepsilon_2^t = 1.6$, $\varepsilon_1^b=2.9$, $\varepsilon_2^b=3.4$.}\label{figpot}
\end{figure}

In fig.~\ref{figpot}, as an illustrative example, the effective potential is shown for a suitable choice of the free
microscopic parameters, in the set-up with $\delta=0$. The minimum is at $\alpha_{min}=9\times 10^{-3}$, corresponding to a
compactification mass $R^{-1} = 8.9$ TeV. The value of the top and bottom quark masses are
$m_t = 176$ GeV and $m_b = 1$ GeV. The Higgs mass is $370$ GeV.

When the bulk-to-boundary couplings are included, the exact spectrum of the bulk-boundary
fermion system defined by the Lagrangian (\ref{Lagferm}) is determined by
solving a complicated transcendental equation (whose form can however be deduced from
eqs.(\ref{boundpott}) and (\ref{boundpotb})).
The lightest states are identified with the top and bottom quarks and are in general
a mixture of localized and bulk fermion states.
When the physical mass induced for the boundary
fields is much smaller than the masses of the bulk fields, to a very good approximation
the top and bottom quark Yukawa couplings (and hence their masses)
are found by integrating out the massive bulk fermions and neglecting the momentum dependence
induced by higher derivative operators.
The relations giving the top and bottom masses are given by appropriate
modifications of eqs.(15)-(18) of \cite{Scrucca:2003ra}.
One has
\begin{equation}
m_a = \left|\frac{m_0^a}{\sqrt{Z_1^a Z_2^a}}\right|\,,\ \ \ a=t,b
\label{mfis}
\end{equation}
where
\begin{eqnarray}
m_0^t \a=\a \frac {\epsilon_1^t \epsilon_2^t}{\sqrt{2} k_t \pi R}\,
{\rm Im}\s f_\delta(\frac{\lambda^t}{k_t},2\alpha)
\label{mu} \;, \nn \\
m_0^b \a=\a \frac {\epsilon_1^b \epsilon_2^b}{k_b \pi R}\,
{\rm Im}\s f_\delta(\frac{\lambda^b}{k_b},\alpha)
\label{md} \;, \nn \\
Z_i^t \a=\a 1
+ \delta_{i1} \frac {\epsilon_1^{b\s 2}\!}{k_b \lambda^b}\,
{\rm Re}\s f_0(\frac{\lambda^b}{k_b} ,0)
+ \delta_{i2}\, \frac {\epsilon_2^{t\s 2}\!}{2k_t \lambda^t}\,
{\rm Re}\s f_0(\frac{\lambda^t}{k_t},0)
+ \frac {\epsilon_i^{t\s 2}\!}{2^{\delta_{i2}}k_t \lambda^t}\,
{\rm Re}\s f_0(\frac{\lambda^t}{k_t},2\alpha)
\label{Ziu} \;, \qquad \nn \\
Z_i^b \a=\a 1
+ \frac {\epsilon_i^{b\s 2}\!}{k_b \lambda^b}\,
{\rm Re}\s f_0(\frac{\lambda^b}{k_b},\alpha)
+ \delta_{i1} \frac {\epsilon_1^{t\s 2}}{k_t \lambda^t}\,
{\rm Re}\s f_0(\frac{\lambda^t}{k_t},\alpha)
\label{Zid} \;.
\end{eqnarray}
The changes induced by the $k_i$'s are better seen in the limit in
which one takes large bulk-to-boundary mixing $\epsilon_1^t,\epsilon_2^t>> 1$
and $\alpha_{min}<<1$. For simplicity, we also take $\epsilon_1^b=\epsilon_2^b=0$, since
we are mainly interested on the top mass formula.
In these approximations one finds
\be
m_{t} \simeq \sqrt{2} k_t m_W F[(2-\delta)\lambda_t/k_t]\,,
\label{mtop-app}
\ee
where
\be
F(x) = \frac{x}{\sinh{x}}\,.
\ee
The function $F(x)$ has a maximum for $x=0$, where $F(0)=1$, and is monotonically decreasing
for $x\geq 0$. Thus
\be
m_{t} \lesssim \sqrt{2} k_t m_W\,,
\label{mtop-bound}
\ee
for both $\delta=0$ and $\delta=1$.
It is clear from eq.(\ref{mtop-bound}) that $k_t\sim 2$ is enough to get the correct top mass.
Another possible way to increase the top mass is obtained by increasing the rank of the
 $SU(3)$ representation
of the bulk fermion which couples to the localized fields. In this way one can get a larger
group-theoretical factor multiplying eq.(\ref{mtop-app}), at the cost of introducing
large representations of SU(3), lowering the Na\"\i ve Dimensional Analysis (NDA) estimate of the cut-off.

\subsection{Estimate of the cut-off}

We estimate the cut-off $\Lambda$ using NDA, as the value at which
the first fundamental coupling in the theory gives rise to one-loop diagrams
of the same size as the tree--level ones.
For simplicity, we consider the non-compact limit $R\rightarrow\infty$, but with
$g_5/\sqrt{2\pi R} = g$ fixed. The 5D loop factor
is $24\pi^3$, so one gets
\be
\frac12\frac{g_5^2\Lambda}{24 \pi^3}=\frac{g^2\Lambda R}{24 \pi^2}\simeq 1\,,
\label{NDAestimate}
\ee
where the factor 1/2 in the first expression  of eq.(\ref{NDAestimate}) is due to the $\Z_2$ orbifold
projection.
We should be careful since $g_5 k$ is effectively a new coupling constant.
The most stringent bounds arise indeed from this coupling, when $g_5$ is the
strong $SU(3)_c$ coupling constant.
One finds $R \Lambda_c \sim 24 \pi^2 /(k g_s^2)\simeq 100/(kR)$, namely that
the cut-off scales
as $1/k$. This rescaling can easily be understood in the non-compact case, by noting that $k$ enters not only
in the coupling, $g_5\rightarrow g_5 k$, but also in the propagators of the virtual states
running in the loop. The latter is reabsorbed by sending $q_5 \rightarrow q_5/k$, $q_5$ being
the momentum along the fifth direction, so that
the loop factor scales as $1/k$.
We see that NDA does not
give strong bounds on the allowed values of $k$, as long as $k\lesssim 10$, which is above
the values we have considered. If one takes instead the electroweak coupling constant,
$\Lambda_w\sim 1000 /(k R)$ and no significant constraint arises.

Although $k$ is practically not constrained by perturbativity, it is important to recall that
the explicit breaking of the SO(4,1) Lorentz symmetry presents the drawback
of generating several counterterms in the effective action which are no
longer constrained by SO(4,1) to be absent or equal between each other.
This would result in a less constrained model and would also lead to the appearance
of additional radiative corrections, absent in the SO(4,1) invariant case.
As an example of an effect of this sort, we would expect that at two-loop level
the Higgs mass will develop a linear divergence.
Indeed, although we think that the Higgs mass term would still be finite,
being associated to non-local operators \cite{Non-local},
the wave function renormalization of the field $A_5$ is no longer exactly cancelled by the
gauge coupling constant renormalization, as in the SO(4,1) invariant case, giving rise to
a divergence for the physical Higgs mass. Since the Higgs mass term is one-loop induced,
such divergence occurs at two-loop level. It is important to stress that this two-loop linear divergence
does not significantly destabilize the Higgs mass.
It would be interesting to better quantify how higher loop corrections, in general,
modify the predictions we have given for the Higgs mass, compactification scale and other parameters.

\section{Results}
\begin{figure}[t]
\begin{center}
\includegraphics*[width=.45\textwidth]{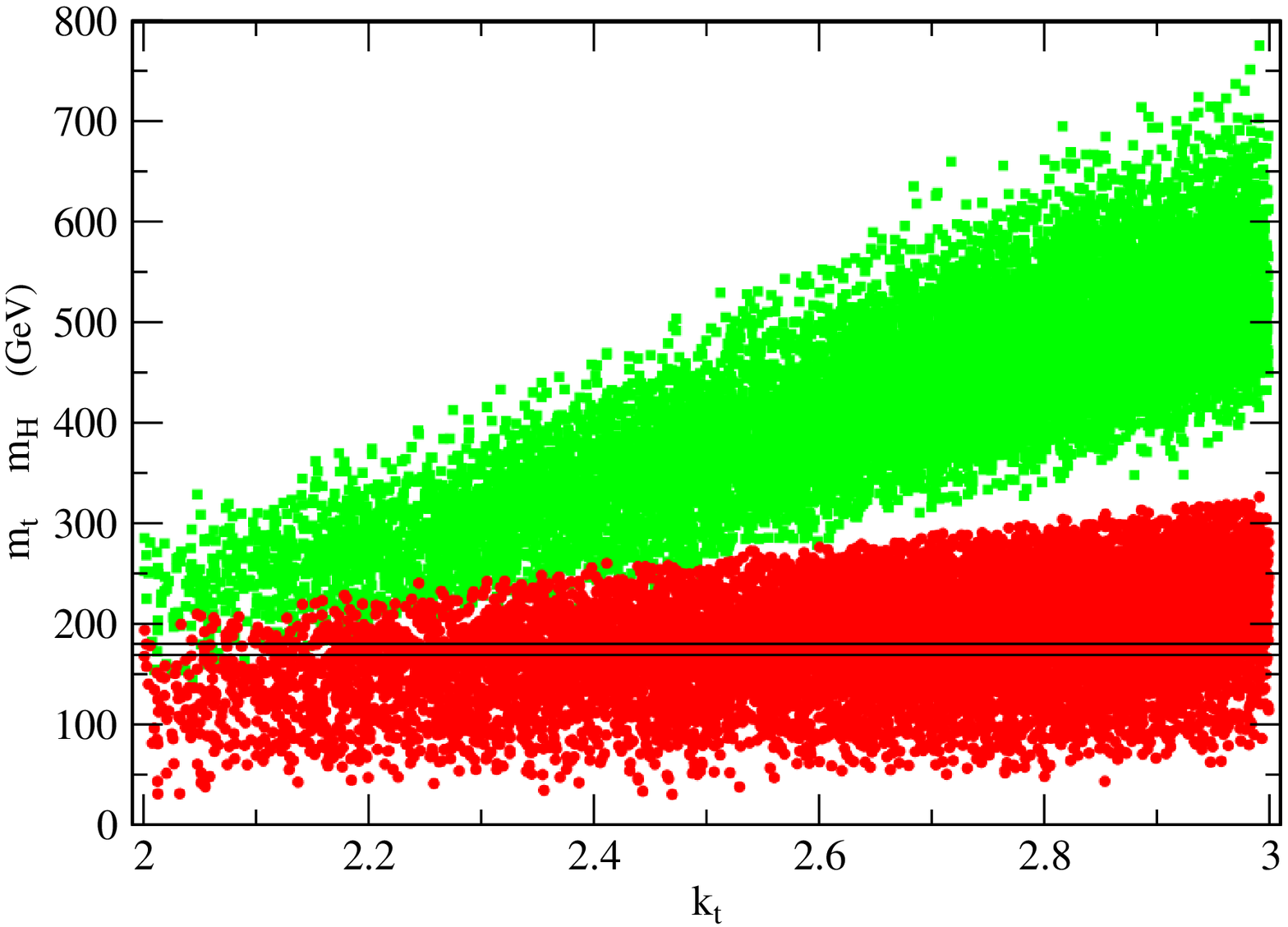}\hspace{5mm}
\includegraphics*[width=.45\textwidth]{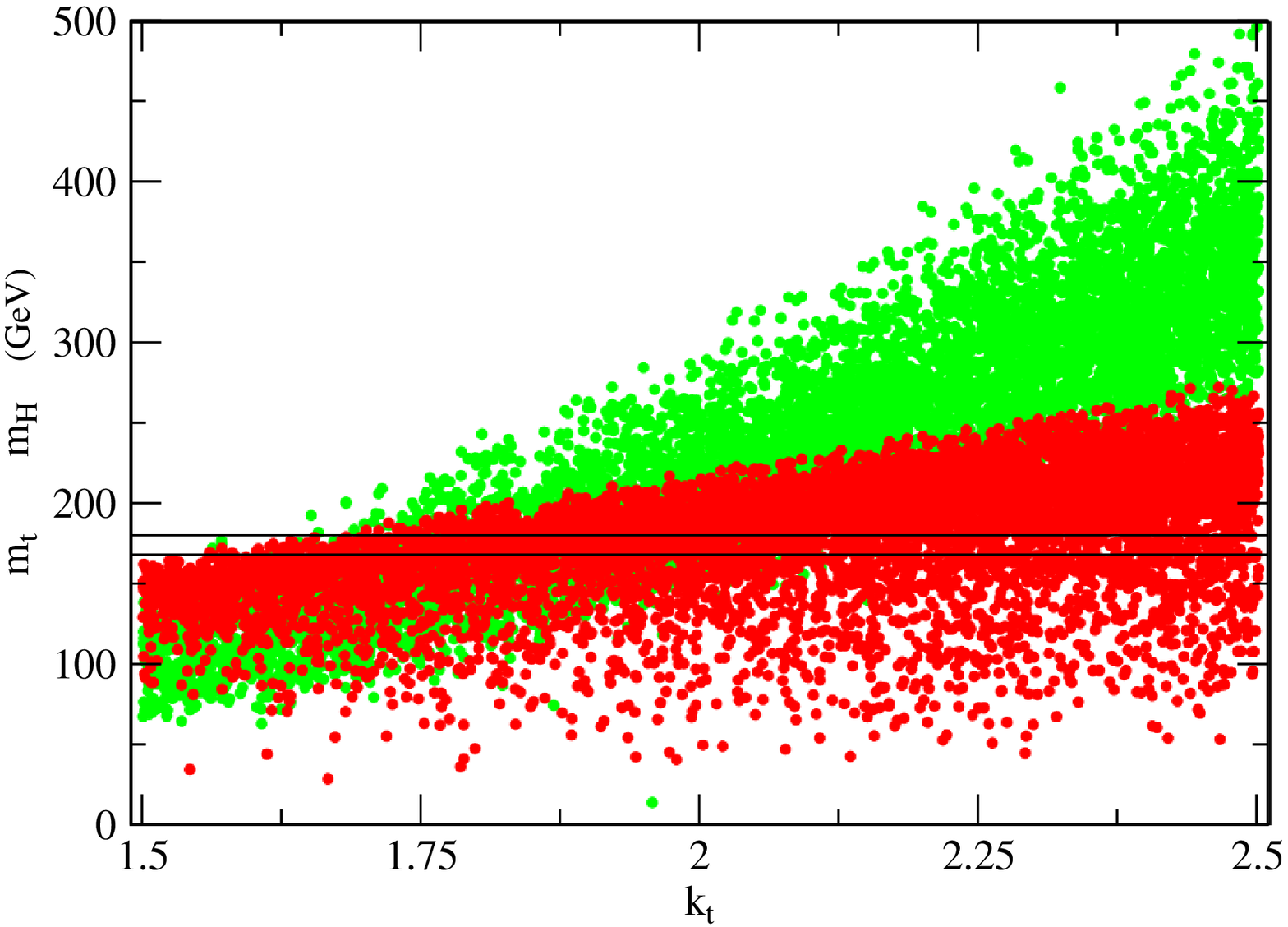}
\caption{Top (red) and Higgs (green) masses versus the input
value of $k_t$ for $\delta=0$ (left) and $\delta=1$ (right). The interval between the two black lines corresponds to the
physical top mass.}\label{fighiggstop}
\end{center}
\end{figure}

In the present section we report the predictions of our model, as obtained by a numerical study.
The analysis is performed by randomly extracting the microscopic parameters
within suitable ranges, and computing the resulting values for the relevant observables,
namely the Higgs and top masses and the compactification scale.
Obviously, we restrict to configurations for which the electroweak symmetry is
 spontaneously broken, so that we discard points for which $\alpha_{min}=0$.
Moreover, a cut $\alpha_{min}<0.05$ is applied in order for the compactification
scale $1/R$ to be sufficiently high.
As described in section~2, we consider two variants of the model ($\delta=0$ and $\delta=1$)
which differ in the location of the boundary fields. The two cases have many qualitative features
in common, but they give rise to different quantitative predictions. In particular,
different ranges are obtained for the Higgs mass.

\subsection{Set-up with $\delta=0$}\label{secsamepoint}

\begin{figure}[tbp]
\begin{center}
\includegraphics*[width=.45\textwidth]{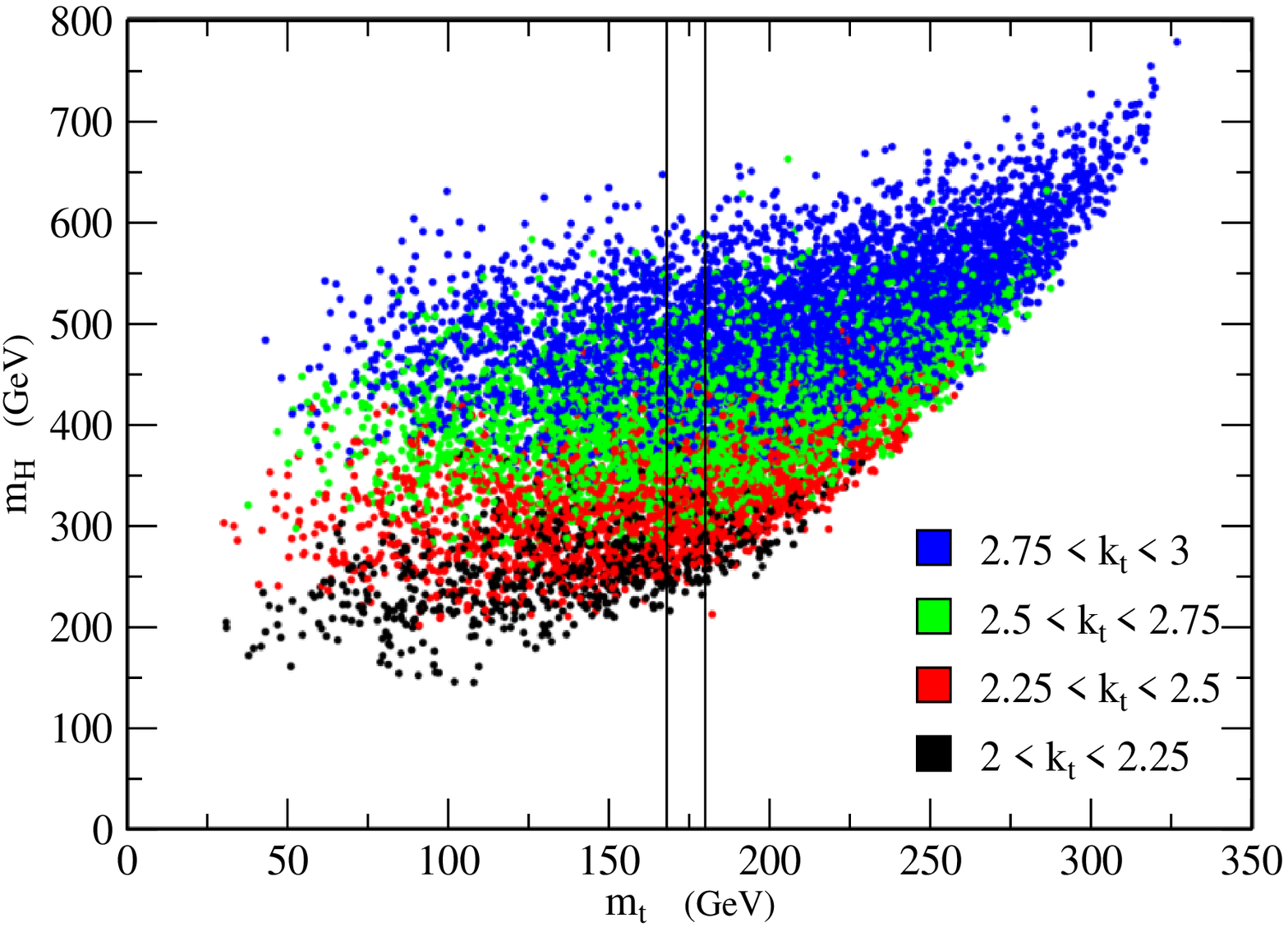}\hspace{5mm}
\includegraphics*[width=.45\textwidth]{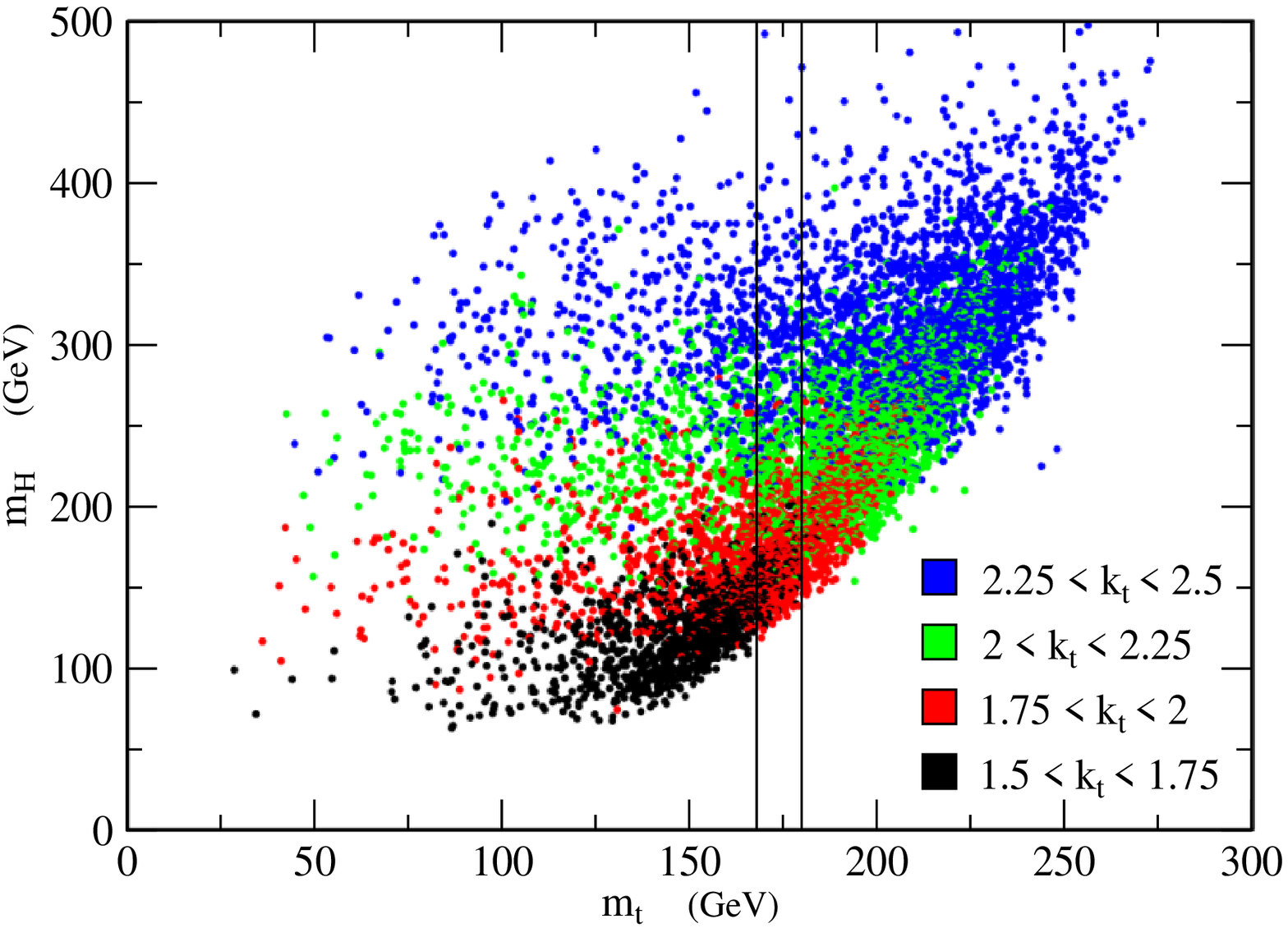}
\caption{Correlation between Higgs and top masses for $\delta=0$ (left) and $\delta=1$ (right).
Different colors label input values of $k_t$ in different ranges and the region
among the two vertical black lines corresponds to the physical top mass.}\label{figmtmh}
\end{center}
\end{figure}

As already mentioned, the cancellation
of the quadratic term in the effective potential, which permits to obtain small
enough values of $\alpha_{min}$, basically results in a correlation between $k_A$
and $k_t$. Our numerical study reveals indeed acceptable points
(with $\alpha_{min}<0.05$) to be only found when $1.1 \times k_t < k_A < 1.5 \times k_t$.
For realizing the plots which follows, the bounds
\begin{displaymath}
\left\{
\begin{tabular}{l}
$2 < k_t < 3$  \\
$2 < k_b < 3$\\
$2.6 < k_A < 3.9$\\
\end{tabular}\right.\,,
\hspace{3em}
\left\{
\begin{tabular}{l}
$0.5 < \lambda^t < 1.5$\\
$5 < \lambda^b < 7$\\
$0.13 < \lambda^A < 3.3$
\end{tabular}\right.\,,
\hspace{3em}
\left\{
\begin{tabular}{l}
$0.75 < \epsilon^t_i < 7.5$\\
$2 < \epsilon^b_i < 7$
\end{tabular}\right.\,,
\end{displaymath}
have been used for the input microscopic parameters.
First of all, in order to appreciate the effect of the Lorentz violating parameters $k_i$, let us see
how the various observables depend on $k_t$.
Clearly, due to the aforementioned cancellation condition,
the behaviour in $k_A$ is similar to the latter, while the results are weakly
sensitive to $k_b$ (and to all others $b$ parameters as well), due to the high value
of $\lambda^b$, which suppresses their contributions.
In figure~\ref{fighiggstop}, the dependence on $k_t$ of the Higgs and top masses is shown.
As expected, the upper bound on the top mass linearly increases
with $k_t$ and correct values (between
the black lines in the figure) are obtained for $k_t\geq 2$.
On the other hand, as expected from eq.(\ref{Va-xCorr}),
the Higgs mass grows quadratically with $k_t$.

\begin{figure}[tbp]
\begin{center}
\includegraphics*[width=.45\textwidth]{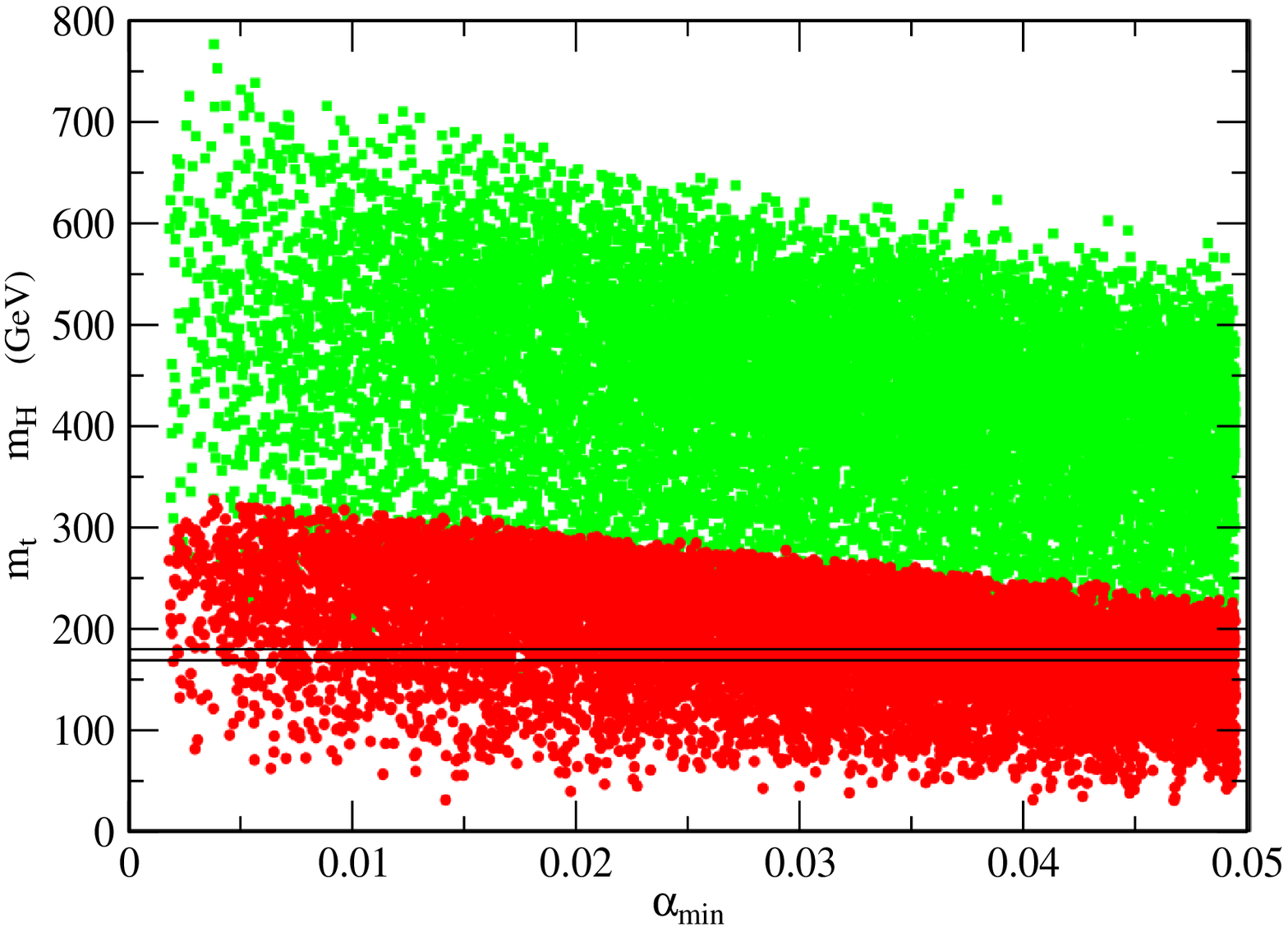}\hspace{5mm}
\includegraphics*[width=.45\textwidth]{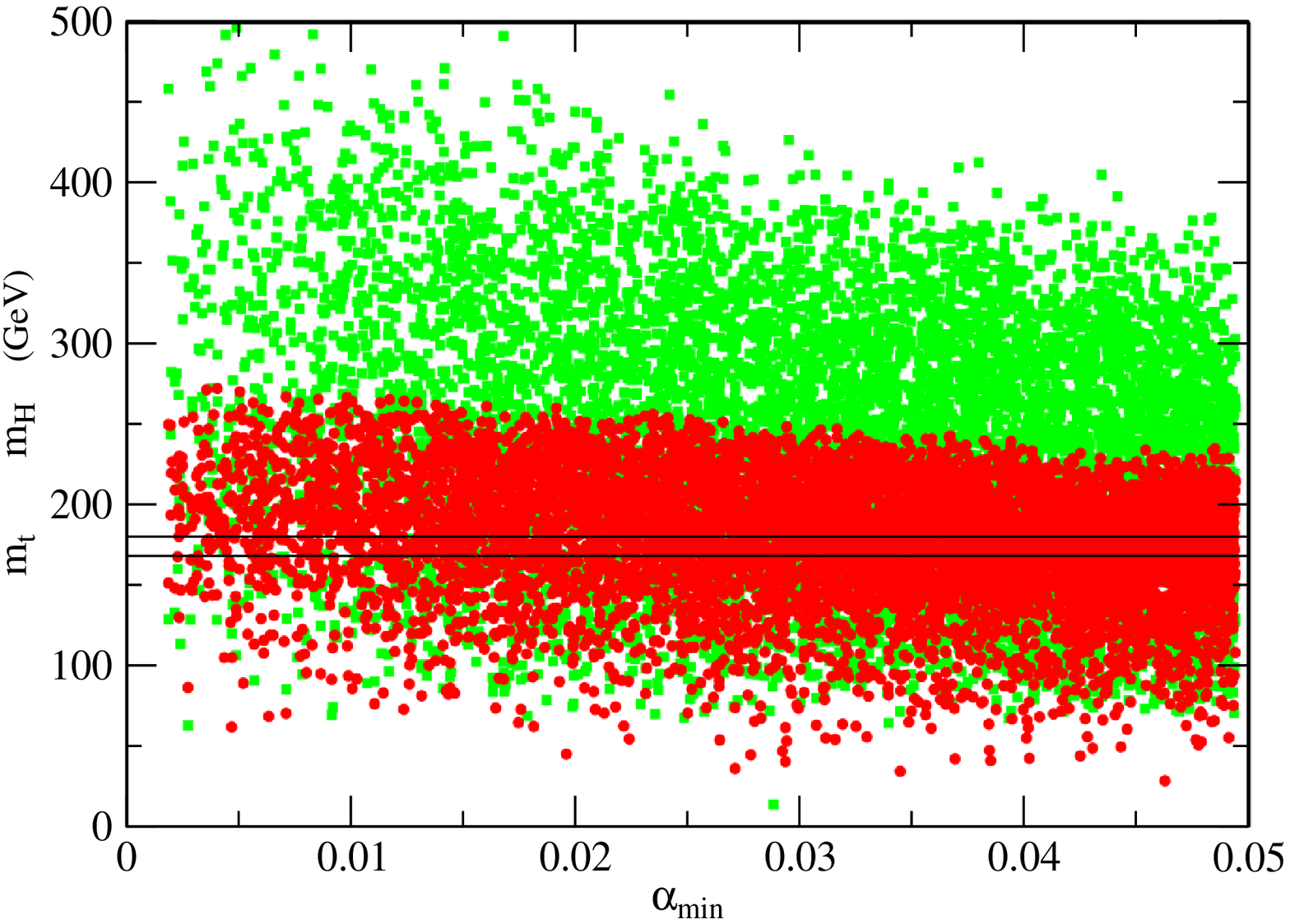}
\caption{Dependence of the Higgs (green) and top (red) masses on $\alpha_{min}$ for $\delta=0$ (left) and $\delta=1$ (right).
The interval between the two black lines corresponds to the
physical top mass.}\label{figtopalpha}
\end{center}
\end{figure}

It can be inferred from figure \ref{fighiggstop} that, at fixed $k_t$, a
certain correlation
between the Higgs and top masses exists. This is shown in fig.~\ref{figmtmh},
in which the Higgs mass is plotted versus the top one, and different colors correspond
to different values of $k_t$.
Figure~\ref{figtopalpha} shows $m_H$ and $m_t$ as a function of $\alpha_{min}$.
Higher Higgs and top masses are favoured
at small values of $\alpha_{min}$, even though realistic values of $m_t$
can always be obtained. The dependence on $\alpha_{min}$ of the upper bound for the top
mass can be derived from eqs.(\ref{mfis}) and (\ref{Zid}) in the large $\epsilon$ regime.

Let us now restrict to realistic values for the top quark mass, in the range
$169\;{\rm GeV} < m_t <180 \; {\rm GeV}$.\footnote{Due to the small statistics of our data, a cut on the bottom mass can not be applied. In the present set of data, $m_b$ goes from $0.2$ GeV up to $10$ GeV,
and is more or less uniformly distributed. As expected, no quantity is found to be correlated with $m_b$, so that realistic values of $m_b$ can be easily obtained.}
With this cut (see fig.~\ref{figmhalphafixmt})
the Higgs mass is found to be in the
range $250-600\ \rm GeV$, independently of the value of $\alpha_{min}$.
In figure~\ref{figlatalphafixmt}, finally, the mass ($M_t = \lambda^t/\pi R$)
of the lightest non-standard fermions in the model, coming from the KK towers of $\Psi_t$ and
$\widetilde \Psi_t$, is plotted versus $\alpha_{min}$.
As we will discuss in section~4,
this mass is important for estimating new physics effects
arising in our model.

\subsection{Set-up with $\delta=1$}

\begin{figure}[tbp]
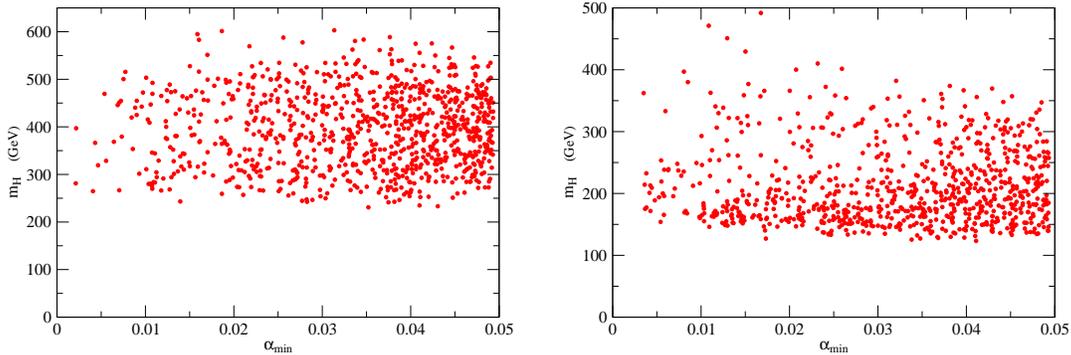

\begin{center}
\includegraphics*[width=.45\textwidth]{mhalphafixmt}\hspace{5mm}
\includegraphics*[width=.45\textwidth]{mhfixmtdelta}
\caption{Higgs mass range as a function of $\alpha_{min}$
for $\delta=0$ (left) and $\delta=1$ (right). The top mass has been fixed to the physical value
$169\ \rm GeV < m_t < 180\ \rm GeV$.}\label{figmhalphafixmt}
\end{center}
\end{figure}

As in the case of the previous subsection,
acceptable vacua are found only if a certain correlation between $k_t$ and $k_A$
is imposed.
We take $1.1 \times k_t < k_A < 1.5 \times k_t$, and we restrict the other
microscopic parameters to the ranges:
\begin{displaymath}
\left\{
\begin{tabular}{l}
$1.5 < k_t < 2.5$  \\
$1.25 < k_b < 2.25$\\
\end{tabular}\right.\,,
\hspace{3em}
\left\{
\begin{tabular}{l}
$0.5 < \lambda^t < 1.5$\\
$5 < \lambda^b < 7$\\
$0.75 < \lambda^A < 3.5$
\end{tabular}\right.\,,
\hspace{3em}
\left\{
\begin{tabular}{l}
$0.75 < \epsilon^t_i < 7.5$\\
$2 < \epsilon^b_i < 7$
\end{tabular}\right.\,.
\end{displaymath}
In figure~\ref{fighiggstop}, the dependence on $k_t$ of the Higgs and top masses is shown.
As in the previous case, the upper bound on the top mass linearly increases with $k_t$,
while the Higgs mass grows quadratically. Note that, differently from the $\delta=0$ case,
configurations with an Higgs mass smaller or equal to the top one can
be found (see also figure~\ref{figmtmh}).
Figure~\ref{figtopalpha} shows the dependence of the
Higgs and top masses on $\alpha_{min}$. The behaviour
is similar to the one for $\delta=0$.

\begin{figure}[tbp]
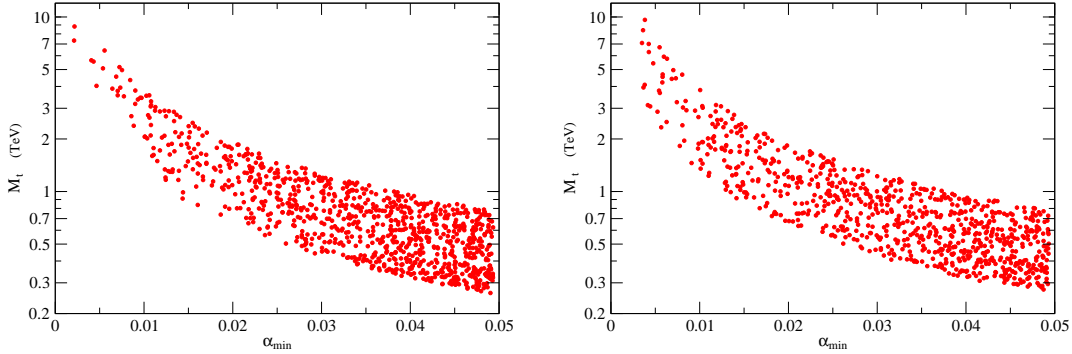

\begin{center}
\includegraphics*[width=.45\textwidth]{Mtalphafixmt}\hspace{5mm}
\includegraphics*[width=.45\textwidth]{Mtfixmtdelta}
\caption{The mass $M_t$ of the first non-standard fermions,
as a function of $\alpha_{min}$ for $\delta=0$ (left) and $\delta=1$ (right).
 The top mass has been fixed to the physical value
$169\ \rm GeV < m_t <\ \rm 180 GeV$.}\label{figlatalphafixmt}
\end{center}
\end{figure}

We now restrict our analysis to configurations with realistic top mass:
$169\ \rm GeV < m_t < 180\ \rm GeV$.
The dependence of the Higgs mass on $\alpha_{min}$
is reported in figure~\ref{figmhalphafixmt}.
Allowed Higgs masses are in the range $125 - 400\ \rm GeV$, and the
distribution favours small values.
Finally, the mass of the lightest non-standard fermions is shown in figure~\ref{figlatalphafixmt}.
The dependence on $\alpha_{min}$ is analogous to the one found
in the case in which all localized fields are at the same fixed point.

\subsection{The EW Phase Transition and a Dark Matter Candidate}

We have also studied the behaviour of our model at finite temperature, focusing in particular
to the study of how (if any) an EW Phase Transition occurs. This analysis is relevant
to establish whether baryogenesis at the electroweak scale could be a viable possibility or not.
As known,
this requires a first-order phase transition where the order parameter $H(T_C)/T_C \geq 1$,
$T_C$ being the critical temperature of the transition.

The analysis is a simple
generalization of \cite{Pan}, so that we will be very brief here
and report only the final results.
The model develops a first-order phase transition at a temperature of order $T_C \sim
(0.1-1.5)/(2\pi R)$. We get $0.01 \leq H(T_C)/T_C \leq 0.05$ for $\delta=0$ and
$0.02 \leq H(T_C)/T_C \leq 0.14$ for $\delta=1$. The phase transition strength,
as expected, is approximately proportional to $1/m_H^2$ and this explains why the
$\delta=1$ set-up appears to have a stronger phase transition than the $\delta=0$ case.
In both cases, however, the latter seems to be too weak to open the possibility
of achieving a baryogenesis at the electroweak phase transition.

Interestingly enough, there is a potential DM candidate particle in our model.
Thanks to the global $U(1)_A$ symmetry that we have imposed to our theory,
under which only the antiperiodic fermions transform, the lowest KK modes of both
$\Psi_A$ and $\widetilde \Psi_A$ are absolutely stable particles.
After EWSB, with the $U(1)^\prime$ assignment we have given, the ${\bf 6}$ of $SU(3)_w$
gives rise to a  couple of four different towers of KK modes (see the appendix of \cite{Scrucca:2003ra} for details),
one for $\Psi_A$ and one for $\widetilde\Psi_A$:
\be
\begin{array}{l@{\hspace{1.5em}}l@{\hspace{1em}}l}
\displaystyle m_n^{(1)}  =  \sqrt{\displaystyle M_A^2+ \frac{k_A^2(n+1/2 +\alpha)^2}{R^2}}\,,& {\rm with} \ \ q=+1\,,&  n\in[ -\infty,+\infty]\,, \\
\rule{0pt}{2.5em}\displaystyle m_n^{(2)}  =  \sqrt{\displaystyle M_A^2+ \frac{k_A^2(n+1/2 + 2 \alpha)^2}{R^2}}\,, & {\rm with} \ \ q=0\,,& n\in[ -\infty,+\infty]\,, \\
\rule{0pt}{2.5em}\displaystyle m_n^{(3)}  =  \sqrt{\displaystyle M_A^2+ \frac{k_A^2(n+1/2)^2}{R^2}}\,, & {\rm with} \ \ q=+2\,,& n\in[ 0,+\infty]\,, \\
\rule{0pt}{2.5em}\displaystyle m_n^{(4)}  =  \sqrt{\displaystyle M_A^2+ \frac{k_A^2(n+1/2)^2}{R^2}}\,, & {\rm with} \ \ q=0\,, & n\in[ 0,+\infty]\,,
\end{array}
\label{DM}
\ee
where $q$ is the electromagnetic $U(1)$ charge of the state.
The lightest particles in eq.(\ref{DM}) are a couple of neutral states with
mass $m_{DM}=\sqrt{M_A^2+k_A^2(1-4\alpha)^2/(4R^2)}$. Since $m_{DM}\geq 1/R$
for the typical input values of $M_A$ and $k_A$, which corresponds to a mass of several TeV,
such states are potential candidates to explain the observed DM abundance in the Universe.

\section{Estimate of Phenomenological Bounds}

In this section we give an order--of--magnitude estimate of the main
 physical effects which we believe to provide the most stringent bounds
on our model. The purpose of this analysis is to show that it
is not trivially ruled out and to roughly estimate the allowed range
in the parameter space of the model.

\subsection{Direct Corrections: the $Z b_L\bar b_L$ vertex and FCNC}

\begin{figure}[tbp]
\begin{center}
\includegraphics[width=.28\textwidth]{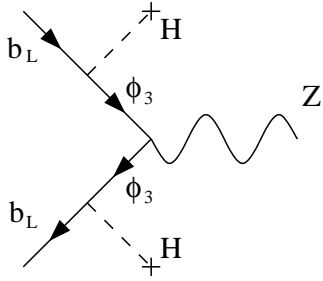}
\caption{The leading tree--level correction to the $Zb_L\bar b_L$
vertex and to the $b_L$ propagator. The distortion of the $Zb_L\bar b_L$ coupling is
generated through the Higgs--mediated mixing of $b_L$ with one component ($\phi_3$)
of the ${\bf 3}_{1/3}$ triplet contained in $\Psi_t$.}
\label{figdiagram}
\end{center}
\end{figure}

The first effect one should worry about is the non--universality of the
EW couplings which is present in our model, after EWSB, since the
physical SM fermions (with diagonal propagators) are a complicated
mixture of fields in different representations of the underlying 5D $SU(3)_w$ EW group.
The $e_{i}^{t,b}$ couplings in Eq.~(\ref{Lagferm}), indeed, make the
localized fields $Q_{L}$ and $b_{R},t_R$ \footnote{The $(t,b)$ couple
should now be thought to represent any of the three families of quarks.}
to mix with the Kaluza--Klein towers of $\Psi_{t,b}$ and $\tl\Psi_{t,b}$
in the ${\bf 6}$ and ${\bf 3}$ representations of $SU(3)$, which contain
singlets, doublets and triplets of $SU(2)_w$. Due to gauge invariance, the
localized fields only couple to the components of the bulk ones with the
right (${\bf 2}_{1/6}$, ${\bf 1}_{-1/3}$ and ${\bf 1}_{2/3}$) quantum
numbers. After EWSB, however, mixing among fields in different representations
are generated. The latter give rise to tree--level corrections to the EW couplings
through tree--level diagrams such as in fig.~\ref{figdiagram}, in which all standard and
non--standard fermions ${q'}_n$, belonging to the Kaluza--Klein towers of
$\Psi$, $\tl\Psi$, propagate. The couplings of ${q'}_n$ to the SM gauge bosons is
 diagonal, since the wave function
of the latter in the extra dimension is flat.
We focus in the following only on the corrections
to the vertex of the $Z$ gauge boson, since this is the one which is experimentally
more constrained.
Diagrams such as
the one depicted in fig.~\ref{figdiagram} give rise at the same time
to a vertex and a propagator correction. The physical correction
to the vertex is obtained only after having canonically normalized
the kinetic terms for the external SM fermions. Gauge invariance allows
Yukawa couplings of the SM Higgs and the bottom quark $b_L$ only through triplets
of $SU(2)$, arising from the ${\bf 6}$ of $SU(3)_w$.
As it is clear from fig.~\ref{figdiagram}, the distortion is inversely proportional to the mass of the triplets
and proportional to the mixing between the doublet and the triplet.
Computing the overlap of the wave functions of the quark doublet with all the
KK tower of the triplets and then diagonalizing the resulting mass matrix is not
straightforward. To a very good approximation, however, the distortion is
dominated by the first massive state of the KK tower, with mass $M_t$.
This is not only the lightest state of the tower, but also the one which mixes
more with the quark doublet. By considering only such a state,
an explicit computation shows that for the down quarks one has
\be
\frac{\delta g_b}{g_b} \simeq \frac{1}{1-\frac 23 \sin^2\theta_W}\frac{\epsilon_1^{t\,2}k_t^2}
{\lambda^{t\,2} Z_1}\Big(\frac{m_W}{M_t}\Big)^2\,,
\label{Zbb}
\ee
where $Z_1$ is the factor appearing in eq.(\ref{Zid}), evaluated at $\alpha=0$, for which
$Z_1=Z_1^t=Z_1^b$. We expect that a similar estimate will also apply for up quarks.
The distortion caused by eq.(\ref{Zbb}) is always safely below current experimental bounds for all light quarks
(including all leptons), in which the bulk fermions are very massive and/or one can consider moderately
small mixing $\epsilon_1^u\lesssim 0.1$.
The only exception is represented by the bottom quark, because the requirement of getting
a reasonable mass for the top quark obliges us to take $\epsilon_1^t\geq 1$ and $\lambda^t\sim 1$.
It turns out, indeed, that eq.(\ref{Zbb}) represents a strong constraint on the parameter
space of the model, as can be seen from figure 12, where $\delta g_b/g_b$ is reported as
a function of $\alpha_{min}$, the most relevant parameter. Considering that $(\delta g_b/g_b)_{exp}\leq
10^{-2}-10^{-3}$, essentially all values of $\alpha_{min}\geq 2 \times 10^{-2}$ ($1/R \lesssim $ 4 TeV)
are ruled out. It is interesting to notice that the constraint imposed by $Zb_L\bar b_L$ also plays
an important role in the warped model of \cite{Agashe:2004rs,Agashe:2005dk}. In the latter case, as in ours,
the requirement of having an acceptable top mass forbids to lower this distortion.

\begin{figure}[tbp]
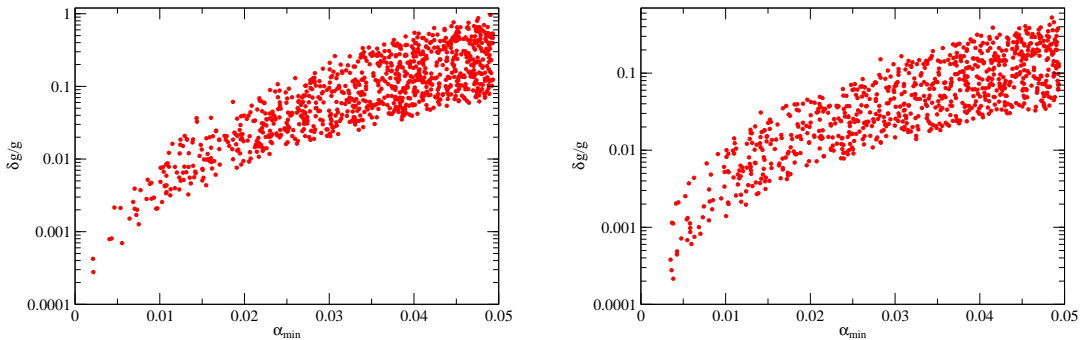

\begin{center}
\includegraphics*[width=.45\textwidth]{ggzero.eps} \hspace{5mm}
\includegraphics*[width=.45\textwidth]{ggdelta.eps}
\caption{Correction to the $Zb_L\bar b_L$ coupling $\delta g_b/g_b$ as a function of $\alpha_{min}$ for
$\delta=0$ (left) and $\delta=1$ (right).}
\label{figggzero}
\end{center}
\end{figure}

Another important issue to consider, closely related to the non--universality of the EW couplings,
is the suppression of the FCNC which are typically generated at tree--level
when integrating out the massive KK modes.
For simplicity, consider here the case in which
the bulk-to-boundary couplings $\epsilon_i$ are diagonal in flavour space
and the non-trivial flavour structure, as in \cite{Scrucca:2003ra}, is totally encoded in non-trivial bulk mass
matrices, which can now involve not only the bulk mass terms $M_i$ but also the Lorentz violating factors $k_i$.
The FCNC are induced in our model by tree-level couplings which arise from diagrams such as the one in fig.~\ref{figdiagram},
in which the non-standard triplet $\phi_3$ switches
at some point to the KK mode of a different family.
Their structure can then be inferred from eq.(\ref{Zbb}), which is in fact
the leading correction to a flavour preserving neutral current or to a FCNC, modulo
the flavour textures which we will not specify here and conservatively take to be of order one.
The typical bound on the couplings of FCNC involving $b$ quarks is $\lesssim 10^{-2}$. Since
these couplings are equal to or smaller than the value estimated by eq.(\ref{Zbb}), they do not
represent any problem. In other words, the strongest bounds in $b$-physics arise from the
$Zb_L\bar b_L$ correction.

The bounds on the couplings of FCNC involving $d$ and $s$ quarks are instead much stronger,
$\lesssim 10^{-5}$. In particular, one should worry that, in presence of a generic flavour mixing,
a light quark ($d$ or $s$) can first switch to a triplet of the heavy
KK tower of the bulk fermion of the corresponding up quark ($u$ or $c$),
which then switches to the much lighter triplet of the KK tower
of the top quark. The latter emits a $Z$ boson and then switches to another heavy
KK tower and thus eventually to another
light quark ($s$ or $d$), resulting in a FCNC. We can estimate the tree--level
coupling $g_{FCNC}$ of this FCNC vertex from eq.(\ref{Zbb}).
Neglecting the factor $Z_1$, which is typically of order 1, one has
\be
g_{FCNC}\sim \frac{\epsilon_1^{c}\epsilon_1^u}
{\lambda^c\lambda^u}\Big(\frac{m_W}{M_t}\Big)^2\,.
\label{FCNC-coup}
\ee
Considering that for the $c$ and the $u$ quarks one can take
$\lambda^c \lambda^u \gtrsim 10$ and, at the same time,
one can naturally take $\epsilon_{1}^{u,c}\sim 0.1$, it is reasonable to expect that
$g_{FCNC}$ can be made smaller than $10^{-5}$ or $10^{-6}$.

FCNC are also induced by the exchange of the massive KK modes of the $Z$ gauge boson and gluons \cite{Delgado:1999sv},
due to the non--universality of their couplings to different families. This effect, which is present even in the absence of EWSB,
comes from the fact that in our model, as required for explaining the mass hierarchy, quarks of different families have different wave functions. By choosing for different generations the same distribution of
localized fields, 
we can however strongly suppress this effect.\footnote{Notice that this is not possible in the set-up
$\delta=1$, which is then disfavored, as far as FCNC are concerned.}
Flavour non--universality in the KK couplings, indeed, arises in this case only from diagrams in which the brane quark $q$ is changed to a bulk KK fermion, which emits a KK gauge boson, and then back to the brane. The effective coupling of this diagram can be estimated as
\be
\frac{\delta g_{KK}}{g_{KK}}\sim\frac{{\epsilon^q}^2}{\lambda^{q\,2}}\,.
\label{gKK}
\ee
For the light families, if $\epsilon^q$ is moderately small ($\lesssim 10^{-1}$), we expect the coupling (\ref{gKK})
to be naturally of order
of $10^{-3}-10^{-4}$, since $\lambda^{q\,2}\gtrsim 10$. In this way, the resulting FCNC --- due to their stronger couplings,
gluons give the dominant contribution ---  is of the same order of magnitude of that estimated for the $Z$ and thus within
the current limits.

\subsection{Oblique Corrections: $\Delta \rho$}

\begin{figure}[tbp]
\begin{center}
\includegraphics[width=.45\textwidth]{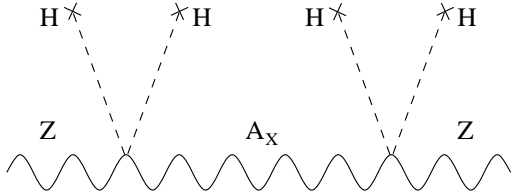}
\caption{Leading tree-level correction to $\Delta \rho$.}
\label{figdeltarho}
\end{center}
\end{figure}

The leading worrisome universal deviation from the SM in our theory is due to the localized mass terms
for the gauge field $A_X$, which are necessarily present, as explained in subsection 2.2,
for anomaly cancellation. They induce tree-level mixing between the $Z$ boson and the
KK modes of $A_X$, which are conveniently expressed in the following term:
\begin{equation}
{\mc L}_m=\frac{1}{2 \cos^2\theta_W}\frac{\alpha^2}{R^2}\Big(Z-\sqrt{3-4\sin^2\theta_W}A_X(y)\Big)^2\,,
\label{hmt}
\end{equation}
where $Z=\cos\theta_W A^3_0-\sin\theta_W A_{Y,0}$ is the usual 4D SM gauge boson, whereas $A_X(y)$ is the
5D field in eq.(\ref{axwf}).
Since, due to the anomaly, the wave functions (\ref{axwf}) are not regular cosines,
the $A_{X,\,n}$ fields couple to the zero--mode of the $Z$ via the $Z$--$A_X$ term in eq.~(\ref{hmt}),
see also fig.~\ref{figdeltarho}.
The physical mass eigenstate of the $Z$, given by eq.(\ref{mzexact}), is then a complicated
mixture of the full KK towers of $A_X$, which lead to a deviation from the SM relation
$m_Z=m_W/\cos\theta_W$ or $\rho=1$. The deviation $\Delta \rho=\rho-1$ (or $T$) is easily computed by
expanding in series the sines in eq.(\ref{mzexact}), or diagrammatically, by considering
the $Z$--$A_X$ mixing terms in eq.(\ref{hmt}). The result is
\be
\Delta \rho = \frac{8\pi^2}{9(1+3\delta)} \alpha_{min}^2 + O(\alpha_{min}^4)\,.
\label{deltarho}
\ee
The current experimental limit on $\Delta \rho$ is
$\Delta\rho_{exp}\lesssim 10^{-3}$, so that eq.(\ref{deltarho}) leads to a bound
on $\alpha_{min}$ which is $\alpha_{min}\lesssim 10^{-2}$ ($1/R\gtrsim 8$ TeV) for $\delta=0$ and
$\alpha_{min}\lesssim 2\times 10^{-2}$ ($1/R\gtrsim 4$ TeV) for $\delta=1$.
For $\delta=0$, eq.(\ref{deltarho}) represents the strongest bound on the model, whereas
for $\delta =1$ it equals the bound found in the last subsection from $Zb\bar b$.

We have not computed the corrections induced by four fermion operators, since we expect that the
associated bounds in our model are roughly the same as the ones estimated in \cite{Barbieri:2004qk}
for universal theories in which the gauge and Higgs field are bulk fields and the SM fermions
are localized states. As such, the resulting bound on $1/R$ is approximately the same
as the one coming from
$Zb_L\bar b_L$ or $\Delta \rho$.\footnote{Notice that in \cite{Barbieri:2004qk} all the oblique parameters,
including the effects of four fermion operators,
are encoded in four parameters, denoted $\hat S$, $\hat T$, $W$ and $Y$. The parameters $\hat S$ and $\hat T$,
modulo a normalization, are defined as in \cite{Peskin:1991sw}, but with respect
to gauge fields which are a mixture of the SM vector bosons with their non standard KK modes.}
Other universal corrections arise at one-loop level.
The leading ones are given by vector-like massive Dirac fermions and are safely small.

\subsection{The Fine--Tuning}

The phenomenological bounds estimated before essentially result in a bound on $\alpha_{min}$
which is $\alpha_{min}\sim 10^{-2}$. As can be seen from, say, figure
\ref{figmhalphafixmt}, low values of $\alpha_{min}$
are not so uncommonly obtained. It is however important to better quantify
how much fine-tuning is necessary to impose in the microscopic parameters of our theory
to get  $\alpha_{min}\sim 10^{-2}$.
The exact determination of such tuning is actually a very challenging task due
to the difficulty of choosing a precise definition of the tuning itself.

The fine-tuning, according to a commonly used definition \cite{Barbieri:1987fn},
is related to the sensitivity of the physical
observables to the microscopic parameters of the theory.
In such a view, the fine-tuning can be estimated by computing the logarithmic
derivative of the observables with respect to the parameters.
In our case the most sensitive observable is the Higgs VEV (namely $\alpha_{min}$), while
the most relevant parameters are the Lorentz violating couplings $k_t$ and $k_A$ or, better,
their ratio $\beta =k_A/k_t$.
Computing the derivative
\begin{equation}
C(\beta) \equiv \frac{\partial \log \alpha_{min}}{\partial \log \beta}\,,
\end{equation}
one finds $C \sim 10^3$ for $\alpha_{min}\sim 10^{-2}$.
If one takes $1/C$ as an estimate of fine-tuning in the model,
this should translate in a fine-tuning of ${\cal O}(\promille)$.

In \cite{Anderson:1994dz}, an improved definition of fine-tuning was proposed.
According to this prescription,
the value of the logarithmic derivative $C(\beta)$ at a given point
must be divided by the average value of $C$ in a suitable range of the microscopic
parameters. This should allow to distinguish ``spurious'' high sensitivity to the
parameters from ``real'' fine-tuning due to cancellations.
In this procedure, however, an appropriate definition of the
range of the microscopic parameters, in which the average will be performed,
must be chosen.
In the present case, the result crucially depends on what we assume to be the ``natural''
values of $\beta$, {\it i.e.} on what we decide to be its ``natural'' interval of variation.
If we take values of $\beta$ for which the EWSB is realized (roughly $0\lesssim \beta\lesssim 1$),
{\it i.e.} all values for which $0<\alpha_{min}<1/2$, and average over all the resulting
 vacua, the fine-tuning turns out to be roughly as before of ${\cal O}(\promille)$.

However, we notice that most of the vacua in this ensemble are either at $\alpha_{min}=0$
or at large values $\alpha_{min}\sim 1/3$. If one disregards these points, by taking
for instance a range of $\beta$ for which $0<\alpha_{min}<1/4$, one finds a relevant
``spurious'' sensitivity of $\alpha_{min}$ on $\beta$.
The range of $\beta$ for which $0<\alpha_{min}<1/4$ is quite small and
the ``real'' fine-tuning is found now, applying the
proposal of \cite{Anderson:1994dz}, to be of ${\cal O}(10\%)$.

Although it is hard to draw a conclusion about the fine-tuning in our model,
in the light of these different estimates, we think that it might be fair
to say that the bound $\alpha_{min}\sim 10^{-2}$
could be translated to a fine-tuning of ${\cal O}(\%)$.

\section{Is Our Model Really a 5D Theory ?}

In this work we have essentially shown how it is possible to get a potentially realistic
model with gauge-Higgs unification at the price of explicitly
breaking the SO(4,1)/SO(3,1) Lorentz generators in the fermionic sector.
In light of this breaking, one could wonder whether it is correct to consider our model
as a ``canonical'' 5D theory or not. Indeed, contrary to the usual ``spontaneous''
breaking of the SO(4,1)/SO(3,1) Lorentz symmetry induced by the compactification, which
implies that at short distances $\Delta x\ll R$ the model is effectively a 5D Lorentz invariant
theory (in the bulk), the explicit breaking we advocate implies that at arbitrarily short scales the SO(4,1)
symmetry is not recovered.
This is clearly a theoretical issue, which is mainly related
to the possible existence and form of an underlying UV completion of our model. Moreover,
the concept of gauge-Higgs unification itself relies on the existence of a 5D interpretation.
It is clear that we can always consider our model as an IR effective description
of a 4D moose theory for which the ``accidental'' SO(4,1) Lorentz symmetry is not
recovered in the fermionic sector \cite{Arkani-Hamed:2001nc}. From this point of view, our model would
resemble more a moose-based little Higgs model rather than a gauge-Higgs unification model.
We would like to point out, however, that the SO(4,1) Lorentz breaking we advocate in this paper
can have a simple origin in the context of a purely 5D theory. A particularly elegant
and interesting explanation is the following.
Consider an axion-like field $\Phi$, which for simplicity we take to be dimensionless,
invariant under the shift $\Phi\rightarrow \Phi+2\pi $.
In light of this shift symmetry,
one can take twisted periodicity conditions for $\Phi$, which reads
\be
\Phi(y+2\pi R) = \Phi(y) + 2\pi \,.
\label{SSshift}
\ee
Scherk-Schwarz reductions of the form (\ref{SSshift}) are not new, appearing in Supergravity
as a way to obtain gauged SUGRA or theories with fluxes (see {\em e.g.} \cite{Bergshoeff:1996ui}).
Consistency of eq.(\ref{SSshift}) with the $\Z_2$ orbifold action $y\rightarrow -y$
requires that $\Phi$ should be $\Z_2$ odd. This is welcome, implying that all the excitations
of $\Phi$ are massive.
Due to the twisted condition (\ref{SSshift}), the
VEV of $\Phi$ is non-trivial. The background configuration $\Phi_0$ which satisfies
the field equations of motion and eq.(\ref{SSshift}) is
\be
\Phi_0(y) = \frac{y}{R}\,,
\label{phi0}
\ee
which clearly induces a spontaneous breaking of the SO(4,1)/SO(3,1) Lorentz symmetry.
The Lorentz violating factors $k$ introduced in eq.(\ref{Lagferm}) are then reinterpreted as due to
couplings involving $\Phi$ and fermion bilinears. It turns out that if we also impose
a $\Z_2$ global symmetry under which $\Phi\rightarrow -\Phi$, the lowest dimensional operator
which couples $\Phi$ and bulk fermions read
\be
\frac{\gamma}{f_\Phi^2}\partial_M \Phi \partial_N \Phi \overline \Psi \gamma^M D^N \Psi\,,
\label{operator}
\ee
where $\gamma$ is a dimensionless coupling and $f_\Phi$ is the ``$\Phi$ decay constant''.
When $\langle\Phi\rangle=\Phi_0$, the operator (\ref{operator}) precisely induces the Lorentz violating
terms which appear in the Lagrangian (\ref{Lagferm}). Since we have considered in our
model values of $k$ which are not close to 1, the effective coupling constant of the operator
(\ref{operator}) is strong (of order 1) and thus insertions of this operator have to be resummed. This is
what we have effectively done in our previous analyses.\footnote{Of course, operators similar to
(\ref{operator}) involving more $\Phi$'s should be taken into account. However, if we assume that
$\gamma$ is large so that one can take $R f_\Phi\ll 1$, these are naturally suppressed.}
Notice that eq.(\ref{phi0}) can also be interpreted as a non-vanishing flux for the 1-form
field-strength $H_1=d \Phi\sim dy$ or else for a non-vanishing flux for the Hodge
dual 4-form field-strength $H_4\sim dx^0\wedge dx^1\wedge dx^2\wedge dx^3$.

We think that the above picture  --- in no way necessary for the model we have presented ---
shows that the Lorentz violating factors $k_j$ can have a natural origin in a 5D
framework.

In light of the rescaling (\ref{rescaling}), the factors $k_j$
effectively imply that different fermions ``see'' a different radius of compactification.
Their effect is then quite similar to recent ideas in the context of Higgless models in 5D warped
models, in which it has been advocated that different fields could
propagate in internal spaces with different sizes \cite{Cacciapaglia:2005pa,Foadi:2005hz}.

\section{Outlook}

In this paper we have shown that realistic models based on gauge-Higgs
unification in 5D flat space can be constructed, but at the price of breaking
the SO(4,1) Lorentz symmetry in the bulk. Our key observation is that the stability of the Higgs
potential is mostly provided by the 5D gauge symmetry rather than the SO(4,1) symmetry.
Breaking the latter results in additional divergencies and in an increasing number of
independent operators to be considered, which however do not significantly
destabilize the Higgs potential.
Somehow, the SO(4,1) breaking models we propose
represent a sort of middle course between little Higgs models and the previously considered
SO(4,1) invariant models with gauge-Higgs unification.
For simplicity, we have focused
our attention on a variant of the minimal model of \cite{Scrucca:2003ra}, where an additional
antiperiodic bulk fermion is introduced. The latter state is crucial to increase the value
of the compactification scale above the TeV scale and, as a by product, its lightest neutral
KK state is a possible DM candidate. Clearly, several other models, already constructed or not,
could be considered in this Lorentz non-invariant scenario.
We have also shown that our model could pass
various phenomenological tests, such as the universality of the couplings,
EWPT and FCNC.

An important issue that we have not considered at all in this paper regards the experimental
signatures of our model. In the light of the forthcoming Large Hadron Collider (LHC), it is
particularly important to address which are (if any) the distinct collider
signatures of our model.
We plan to address in a future work the latter issue, as well as a detailed study of the viability
of our theory as a realistic proposal to go beyond the SM.

\section*{Acknowledgments}

We would like to thank G. Cacciapaglia, C. Csaki, A. Pomarol, A. Romanino, C.A. Scrucca, L. Silvestrini, A. Strumia
and  P. Ullio for useful discussions and comments.
We also thank G.~Cacciapaglia, C.~Csaki and S.C.~Park for sharing a draft of
 \cite{csaba-giacomo} with us prior to publication.
This work is partially supported by the European Community's Human
Potential Programme under contract MRTN-CT-2004-005104 and by the Italian
MIUR under contract PRIN-2003023852.

\end{document}